\documentclass[aps,showpacs,amsmath,amssymbd,dvips,showkeys,linenumbers,preprint,preprintnumbers]{revtex4}
\usepackage{graphicx}
\usepackage{color}

\def \be  {\begin{equation}}
\def \ee  {\end{equation}}
\def \ee  {\end{equation}}
\def \bea {\begin{eqnarray}}
\def \eea {\end{eqnarray}}

\newcommand{\nn}{\nonumber}

\begin{document}

\preprint{ECTP-2017-03}
\preprint{WLCAPP-2017-03}
\hspace*{4mm}

\title{Quark-hadron phase structure, thermodynamics and magnetization of QCD matter}

\author{Abdel Nasser Tawfik}
\email{atawfik@rcf.rhic.bnl.gov}
\affiliation{Egyptian Center for Theoretical Physics (ECTP), Modern University for Technology and Information (MTI), 11571 Cairo, Egypt}
\affiliation{World Laboratory for Cosmology And Particle Physics (WLCAPP), Cairo, Egypt}

\author{Abdel Magied Diab} 
\email{a.diab@eng.mti.edu.eg}
\affiliation{Egyptian Center for Theoretical Physics (ECTP), Modern University for Technology and Information (MTI), 11571 Cairo, Egypt}
\affiliation{World Laboratory for Cosmology And Particle Physics (WLCAPP), Cairo, Egypt}

\author{M. T. Hussein}
\email{tarek@sci.cu.edu.eg}
\affiliation{Physics Department, Faculty of Science,  Cairo University, 12613 Giza, Egypt}

\begin{abstract}

SU($3$) Polyakov linear-sigma model (PLSM) is systematically implemented to characterize the quark-hadron phase structure and to determine various thermodynamic quantities and magnetization of the QCD matter. In mean-field approximation, the dependence of the chiral order-parameter on finite magnetic field is also calculated. In a wide range of temperatures and magnetic field strengths, various thermodynamic quantities including trace anomaly, speed of sound squared, entropy density, specific heat are presented and some magnetic properties are described, as well. Wherever available these results are confronted to recent lattice QCD calculations. The temperature dependence of these quantities confirms our previous result that the transition temperature is reduced with the increase in the magnetic field strength, i.e. QCD matter is to be characterized by an inverse magnetic catalysis. Furthermore, the temperature dependence of the magnetization shows that the conclusion that the QCD matter has paramagnetic properties slightly below and far above the pseudo-critical temperature, is confirmed, as well. The excellent agreement with recent lattice calculations proves that our QCD-like approach (PLSM) seems to possess the correct degrees-of-freedom in both hadronic and partonic phases and describes well the dynamics deriving confined hadrons to deconfined quark-gluon plasma. 

\end{abstract}

\pacs{21.65.Mn, 12.39.Fe, 25.75.Nq, 12.38.Mh, 74.25.N}
\keywords{Equations of state in nuclear matter, Chiral Lagrangian, Quark confinement, Quark-gluon plasma, Response to electromagnetic fields}

\maketitle
\tableofcontents
\makeatletter
\let\toc@pre\relax
\let\toc@post\relax                 
\makeatother 

\section{Introduction \label{Intro}}

The phase structure of the quantum chromodynamic (QCD) matter at very high temperatures [much higher than the pseudo-critical temperature ($T_c$), which characterizes the hadron-quark deconfinement phase transition] is well understood by the perturbation theory, for instance, that close to the pseudo-critical temperature and especially at high temperatures \cite{Kajantie:2003AS, Andersen:2011QD} tools describing the equation of  state (EoS) are well developed. The perturbative EoS of the QCD matter and their implications to neutron star physics were discussed  \cite{Vuorinen2016}. Three-loop corrections including the effects of finite quark masses were introduced  \cite{Kurkela2010GH}.  Furthermore, the perturbative pressure has been calculated to different leading orders \cite{Rummukainen10, Rummukainen11, Shuryak1978, Kapusta1979, Toimela1983, Arnold1995}. The current state-of-the-art was briefly discussed in Ref. \cite{Freedman1978AA, Vuorinen2003}. 

On one hand, the entire quark-hadron phase structure still represents a great challenge to lattice and particle scientists. On the other hand, it is difficult to draw predictions for the properties of the QCD matter, especially the partonic matter, at arbitrary finite temperatures and chemical potentials. The non-perturbative approaches, such as the lattice QCD simulations, are the only reliable first-principle tools across the entire phase-transition. The perturbation theory gives another approach at high temperatures. This predicts that crossover to quark-gluon plasma (QGP) occurs at $T\sim{\cal O}(m_{\pi})$ and might be tolerated to be implemented at finite chemical potentials much less than the temperatures. However, this approach can't be used at larger chemical potentials and lower temperatures, i.e. color-superconducting phase structure, for instance. The reason is that the QCD matter under such extreme conditions of temperatures and densities remains strongly coupled. In the present work, we shall observe how an additional extreme condition, the magnetic field, greatly influence the quark-hadron phase structure. 

We first recall the origin of such a magnatic field and why it shouldn't be ignored? In off-central heavy-ion collisions (HIC) and due to oppositely relativistic motion of electric charges, a huge magnetic field can be created. But because of the very short lasting time-span (life-time) of the evolution of such a magnetic field, it is assumed to have almost no effect on the detector and on its external magnet but remarkable influences on the strongly interacting QCD matter. In addition to hot and dense QCD matter, the influence of a strong magnetic field ($e B$) on the quark-hadron phase structure is also a very interesting subject. This study has a realistic strong relevance to the phenomenology of the relativistic heavy-ion collisions in which a strong magnetic field is likely produced especially in non-central collisions \cite{Kharzeev2008, Skokov:2009}. It is conjectured to have essential effects at the early stages of HIC, where the response of the magnetic effect is assumed to have a large medium-effect. The latter depends on the variation of the magnetic diffusion time \cite{Synchrotron:2010, Elec:Magnet} and the electrical conductivity due to the effect of magnetic field on the medium itself \cite{Gupta:2004, Bratkovskaya:1995}. On the other hand, this study is related to important {\it final-state}  phenomenon characterizing the QCD matter, such as chiral magnetic effect (CME) and magnetic catalysis \cite{Shovkovy:2013,Preis:2011,Sanfilippo:2010,catalysis:2014,Catalysis:2015}. CME is strongly sensitive to the electric charge separation phenomenon that probably can be measured in HIC experiments such as ALICE at LHC \cite{ALICE:2012}, PHENIX \cite{PHENIX:2014} and STAR \cite{STAR:2009,STAR:2010,STAR:2011,STAR:2014} at RHIC. In different HIC experiments such as superproton synchrotron (SPS), relativistic heavy-ion collider (RHIC) and large hadron collider (LHC), tremendously huge magnetic fields are expected, e.g. they range between $0.1\, m_{\pi}^2$ to $m_{\pi}^2 $ and $10-15\, m_{\pi}^2$, respectively \cite{Skokov:2009, Elec:Magnet}, where $m_{\pi}^2\simeq 10^{19}$ Gauss. 
 
So far, there are various numerical calculations of the experimentally measured effects of the magnetic fields, such as the lattice QCD simulations \cite{Fodor_eB:2012, lattice:2013b, lattice:2013, Endrodi:2013, QCD:2013d, LQCD:Magnet2014}, the QCD phase diagram in an external magnetic field \cite{Fodor_eB:2012,Mizher;2010, Skokov;2012}, squeezing QCD matter \cite{squeezing:2013}, hadron resonance gas (HRG) model \cite{HRG1,HRG2}, SU($2$) NJL \cite{Klevansky:1992, NJLsu2} and NJL with Ployakov loop corrections \cite{Menezes:2009a, Fukushima:2010l}, and PLSM  \cite{Mizher;2010, Skokov;2012, Ruggieri:2013}. Moreover, different theoretical models reveal interesting features about the response of hot and dense medium to finite the magnetic field strength, such as the higher-order moments of the quark multiplicity \cite{Tawfik:LSM}, the chiral phase structure of various meson states in hot and dense medium \cite{Tawfik:Masses} and in SU(3) PLSM  at finite magnetic field \cite{Tawfik:Magnetic, Tawfik:Magnetic2}, the thermodynamics and higher-order moments \cite{Tawfik:Magnetic}, the electromagnetic properties of thermal  QCD matter at vanishing and finite magnetic fields \cite{Diab1455, Diab2355, Diab23}.  Great details on understanding the phase structure of strongly interacting QCD matter in strong magnetic fields are reviewed in Refs. \cite{Shovkovy:2013,review1,review2,review3}. 
 
Now we characterize shortly the approach which shall be utilized; the (non)linear-sigma model \cite{lsm1}. In this QCD-like model, the spinless sigma meson, the scalar which was introduced by Schwinger in 1957 \cite{Schwinger}, accompanies the triplet $\pi$ mesons, where the elementary particles could be described through the theory of quantized fields \cite{Schwinger2}. Interesting readers are kindly advised to consult Refs. \cite{Fukushima:2010l, Ruggieri:2013,Tawfik:Magnetic,Tawfik:Magnetic2,review1,review2, review3, THD:magnetic}. These can be compared with studies conducted at vanishing magnetic fields \cite{Tawfik:LSM, Tawfik:Masses, Tawfik:quasi, Tawfik:2013eua}. Many of these studies have investigated the quark-hadron phase transition in thermal and dense medium and thus likelt extend the QCD phase-diagram. Following this procedure, we performed our calculations for the quark-hadron phase structure and the QCD EoS from the PLSM. Furthermore, we compare our results to the available lattice QCD calculations at zero \cite{lQCD2014} and nonzero magnetic fields \cite{LQCD:Magnet2014}. 

With this regard, it was pointed out that the Large Hadron Collider (LHC) at CERN seems to detect possible signatures for the CP violation and the chiral symmetry restoration \cite{fraga2008, fraga2009}. Also, the QCD vacuum properties have been studied by means of the holographic QCD models \cite{holographicMS}. Empirical detection for CME was described in Ref. \cite{Kharzeev2008, Fukushima:2008}. The impacts of finite magnetic field strength have been investigated within different frameworks, mainly, by using effective models \cite{Klevansky:1989, Gusynin:1995, Semenoff:1990, Goyal:2000, Hiller:2008, Rojas:2008}, for instance, NJL model \cite{Fukushima:2010l,Rojas:2008, Klevansky:1992, Klimenko:1998, Babansky:1998, Menezes:2009a}, PLSM \cite{Tawfik:Magnetic, Tawfik:Magnetic2, THD:magnetic} and the chiral perturbation theory ($\chi$PT) \cite{Fukushima:2010l,Cohen:Cohen2007, Agasian:2000}. These are examples on the phenomenologically interesting consequences of finite magnetic field strength on the QCD matter. As mentioned, by applying finite magnetic field strength on the QCD matter important phenomenon such as CME and the magnetic catalysis \cite{Shovkovy:2013,Preis:2011,Sanfilippo:2010,catalysis:2014,Catalysis:2015} can be described. CME is strongly related to the electric charge separation phenomenon that probably become measurable in HIC experiments such as PHENIX \cite{PHENIX:2014} and STAR \cite{STAR:2009, STAR:2010, STAR:2011, STAR:2014} at RHIC and ALICE at LHC \cite{ALICE:2012}. The magnetic catalysis characterizes the dependence of $T_c$ on the magnetic field strength. 

In the present work, we introduce a systematic study aiming at characterizing the respective influences of  magnetic field on the quark-hadron phase transition in a wide range of temperatures. We utilize SU($3$) PLSM in mean field approximation. Prior to confronting our results to recent lattice QCD calculations, we shall show that the order parameters have reasonable agreements with the lattice predictions. Whenever available, we shall compare our calculations for various thermodynamic quantities with different lattice QCD calculations. The characterizations of the hadronic EoS have been discussed in Ref. \cite{satz2009, Tawfik:2012ty}. It was concluded that the characteristics of the speed of sound squared, $c_s^2 =\partial p/\partial \epsilon|_{s/n}\equiv s/c_v$ and the $p/\epsilon$ ratio \cite{Tawfik:2012ty}. Both quantities are assumed being distinguishable below and above $T_c$, while below $T_c$, the thermal behavior of the speed of sound squared was found matching well the $p/\epsilon$ ratio.  Above $T_c$, the value of $c_s^2=s/c_v$ becomes larger than that of $p/\epsilon$.  At very high temperatures, both quantities get very close to each other and likely approach the asymptotic value, i.e. $1/3$  \cite{lqcd4}. We shall calculate the chiral order-parameter ($M_b$), and the light and strange quarks net-condensate ($\Delta_{l,s}$) at finite temperature and magnetic field strength and vanishing chemical potential. The temperature dependence of various thermodynamic quantities and some magnetic properties of the QCD matter shall be calculations. The influences of finite magnetic field and the Landau level quantization on the chiral quark-hadron phase transitions shall be analyzed. We present the temperature dependence of the PLSM order parameters and the subtracted chiral condensates. Furthermore, we shall study these effects on the thermodynamic quantities including trace anomaly, speed of sound squared, entropy, and specific heat, etc. The magnetic properties of the thermal QCD matter shall be described through the magnetization and the magnetic catalysis. Our calculations are confronting to recent lattice QCD simulations. Accordingly, the thermal QCD matter seems to have paramagnetic properties and inverse magnetic catalysis.

The present paper is organized as follows. Section \ref{sec:model} introduces the set-up of the SU($3$) PLSM in thermal and dense QCD medium at zero and nonzero magnetic field. The quark-hadron phase structure shall be given in section \ref{structure}. The order parameters of the chiral condensates and the deconfinement phase transitions at vanishing and finite magnetic field shall be elaborated in section \ref{prder_parameter}. The subtracted condensates are confronted to recent lattice QCD simulations, in section \ref{Sub_condensates}. Various PLSM thermodynamic quantities are also compared with recent lattice QCD calculations at vanishing and finite magnetic field strength in section \ref{thermo}. Last but not the least, the magnetic properties of the thermal QCD matter such as the magnetization and the magnetic catalysis shall be presented in section \ref{Mag_properties}.  Section \ref{Conclusion} is devoted to the final conclusions.

\section{SU(3) Polyakov Linear-sigma Model \label{sec:model}} 

The linear-sigma model is a widely acceptable approach for the investigation of meson states \cite{Tawfik:Masses}. The chiral SU($3$) LSM Lagrangian with $N_f=2+1$, e.g. two light and one heavy strange quarks, consists of two parts; $\mathcal{L}_{chiral}=\mathcal{L}_q+\mathcal{L}_m$.
\begin{itemize}
\item The first term expresses the quark contributions, Eq. (\ref{lfermion}), which are coupled with a flavor-blind Yukawa coupling ($g$) \cite{blind}, i.e. quarks couple to mesons, 
\begin{eqnarray}
\mathcal{L}_q &=& \sum_f \overline{q}_f(i\gamma^{\mu}
D_{\mu}-gT_a(\sigma_a+i \gamma_5 \pi_a))q, \label{lfermion} 
\eea
which isn't depending on the quark current masses and thus shouldn't contribute to the quark condensates, directly,  
\item while the second  term gives the meson contributions
\begin{eqnarray}
\mathcal{L}_m&=&\mathrm{Tr}(\partial_{\mu}\Phi^{\dag}\partial^{\mu}\Phi-m^2
\Phi^{\dag} \Phi)-\lambda_1 [\mathrm{Tr}(\Phi^{\dag} \Phi)]^2 -
\lambda_2 \mathrm{Tr}(\Phi^{\dag}
\Phi)^2 \nn \\ &+& c[\mathrm{Det}(\Phi)+\mathrm{Det}(\Phi^{\dag})]
+\mathrm{Tr}[H(\Phi+\Phi^{\dag})],  \label{lmeson}
\end{eqnarray}
where $\Phi$ is $3\times3$ matrix includes the nonet meson states 
\bea
\Phi &=&  \sum_{a=0}^{N_f^2 - 1} T_a (\sigma_a -i \pi_a).
\eea 
The number of generators ($T_a$) is defined according to the number of the quark flavors ($N_f$).  At $N_f=3$, i.e. three degenerate quark flavors, the chiral Lagrangian contains unitary matrices with  SU$(3)_r\times$SU$(3)_l$ symmetry \cite{Weinberg1972}. In  U$(3)$ algebra, $T_a$ is determined by Gell-Mann matrices $\hat{\lambda}_a$ \cite{Weinberg1972}; $T_a= \hat{\lambda}_a/2$ with $a=0,\cdots, 8$. 
\end{itemize}

In order to incorporate the deconfinement phase-transition in such a QCD system, LSM can be coupled to the Polyakov-loop potential ($\mathbf{\mathcal{U}}(\phi, \phi^*, T)$) representing the degree(s) of freedom and the gluons dynamics. Its functional form is motivated the QCD symmetries in the pure gauge limit. This potential should be fine-tuned through matching with recent lattice QCD simulations, for instance. Different expressions are fitted to pure gauge lattice so that various thermodynamic quantities can be well reproduced. This adds color-gluon interaction such that $\mathcal{L}=\mathcal{L}_{chiral}-\mathbf{\mathcal{U}}(\phi, \phi^*, T)$. Furthermore, $Z(3)$ symmetry \cite{Schaffner:2013 chiral, Ratti:2005, Roessner:2007, Fukushima:2008} should be guaranteed. For example, in the pure gauge limit, a temperature-dependent potential can be used; $U(\phi, \phi^{*},T)$. Its $Z(3)$ center symmetry should be similar to that of the pure gauge QCD Lagrangian \cite{Ratti:2005,Schaefer:2007d}. Trace of the color space of the Polyakov loop determines the creation operator of a static quark at the spatial position $\vec{x}$. The color charges and the gluons dynamics can then be given in dependence on the thermal expectation value of a color-traced Wilson-loop in the temporal direction \cite{Polyakov:1978vu},
\bea
\phi = \langle\mathrm{Tr}_c\, \mathcal{P}(\vec{x})\rangle/N_c, \qquad && \qquad
\phi^* = \langle\mathrm{Tr}_c\, \mathcal{P}^{\dag}(\vec{x})\rangle/N_c. \label{phis}
\eea
Then, the Polyakov-loop operator becomes identical to the Wegner-Wilson loop \cite{Polyakov:1978vu},
\begin{eqnarray}
 \mathcal{P}(\vec{x})=\mathcal{P}\; \exp{\Big[i\int_0^{1/T}d \tau A_0(\vec{x}, \tau)\Big]}, \label{loop}
\end{eqnarray}
where $A_0(\vec{x}, \tau)$ is the temporal component of the Euclidean gauge field $A_\mu$ \cite{Polyakov:1978vu, Susskind:1979up} and $\tau$ denotes the Euclidean time component. It was found that enlarging $N_c$ decreases the pseudo-critical temperature of the deconfinement phase-transition ($T_c$) \cite{Tawfik:Masses}. Thus, the Polyakov loops $\phi$ and $\phi^{*}$ can be considered as order parameters for the deconfinement phase-transition \cite{Ratti:2005,Schaefer:2007d}. 

There are different potential types for the Polyakov loop corrections \cite{Ratti:2005, Schaefer:2007d, Roessner:2007, Fukushima:2008}. In the present work, we implement the polynomial form, where the  parameters were estimated within the perturbation theory \cite{Banks1982, Miransky1997, Appelquist1996, Braun2006}. This characterized an $N_f$-dependent decrease of the critical temperature ($T_0$), while other potential forms fix their parameters to be compatible with recent lattice calculations. 
 \begin{eqnarray}
\frac{\mathbf{\mathcal{U}_{poly}}\left(\phi, \phi^*, T\right)}{T^4}=-\frac{b_2(T)}{2}\left(\left|\phi\right|^2 + \left|\phi ^*\right|^2\right)-\frac{b_3
}{6}\left(\phi^3+\phi^{*3}\right)+\frac{b_4}{16}\left(\left|\phi\right|^2 +\left|\phi^*\right|^2\right)^2, \label{Uloop}
\end{eqnarray}
where $b_2(T)=a_0+a_1\left(T_0/T\right)+a_2\left(T_0/T\right)^2+a_3\left(T_0/T\right)^3$. With  $a_0=6. 75$, $a_1=-1. 95$,  $a_2=2. 625$,  $a_3=-7. 44$, $b_3 = 0.75$ and $b_4=7.5$ \cite{Ratti:2005}, the pure gauge QCD thermodynamics is well reproduced. For a better agreement with recent lattice QCD simulations, the critical temperature $T_0$ should be fixed to $187~$MeV for $N_f=2+1$ \cite{Schaefer:2007d}. The possible differences in $T_0$ as used in PNJL model and  PLSM, $187$ and $220~$MeV, respectively, can be understood due to the introduction of quarks which in the pure gauge limit leads to a decrease in the deconfinement temperature. In both models, the Polyakov loop potential ($U$) should be the same.

\begin{table}[htb]
\begin{center}
\begin{tabular}{|c | c | c | c | c | c | c |}
\hline
$m_\sigma$ [MeV] & $c\,$ [MeV] & $h_l\,$ [MeV$^3$] & $h_s\,$ [MeV$^3$] & $m^2 \,$ [MeV$^2$] & $\lambda _1$ & $\lambda _2$\\ 
\hline 
800 & $4807.84$ & $(120.73)^3$ & $(336.41)^3$ & -$(306.26)^2$ & $13.49$& $46.48$\\ 
\hline 
\end{tabular}
\caption{Summary of various PLSM parameters.  A detailed description is given in Ref. \cite{Schaefer:2008hk}  \label{tab:1a}}
\end{center}
\end{table} 

In thermal equilibrium, the exchanges of energy between particles and antiparticles can be described by path integral over quark, antiquark and meson fields. Thus, the grand canonical partition function reads   
\begin{eqnarray}
\mathcal{Z}
&=& \int\prod_a \mathcal{D} \sigma_a \mathcal{D} \pi_a \int
\mathcal{D}q \mathcal{D}\bar{q} \;\exp{ \left[\int_x
\mathcal{L}^E \right]}, \label{partitionFunc}
\end{eqnarray}
where $\int_x\equiv i \int^{1/T}_0 dt \int_V d^3x$ with $V$ being the volume of the system and $\mu_f$ is the chemical potential for quark flavors $f=(u, d, s)$. For SU($3$) quark flavors, $\mathcal{L}^E$ is the Euclidean Lagrangian. The three quark chemical potentials can be included in  
\bea
\mathcal{L}^E = \mathcal{L}_{chiral} +\sum_{f=u, d, s} \mu_f \bar{q}_f \gamma^0 q_f 
\eea
As highlighted, we assume symmetric quark matter and degenerate light quarks. Thus, a uniform blind chemical potential can be defined $\mu_f \equiv \mu_{u, d} = \mu_s$ \cite{blind, Schaefer:2007c, Schaefer:2008hk}. 

In mean-field approximation, the calculations of the partition function is conducted in a similar procedure as elaborated in Ref. \cite{blind, Schaefer:2007c}. Accordingly, we replace the meson fields in the exponent in Eq. (\ref{partitionFunc}) by their expectation values. In doing this, the  quantum and thermal fluctuations of the mesons are neglected while the quarks and antiquarks are retained as quantum fields \cite{blind, Schaefer:2007c, Schaefer:2008hk}. The integration over the fermions can be written as a trace over a logarithm and the trace should be evaluated in Matsubara formalism \cite{Kapusta:2006pm}. On the other hand, the mesonic fields in Eq. (\ref{partitionFunc}) are determined by their nonvanishing vacuum expectation values $\phi= T_0 \bar{ \sigma_0}+T_8\bar{{\sigma_8}}$. More details can be found in Refs. \cite{blind, Schaefer:2007c, Schaefer:2008hk}. Finally, all thermodynamical quantities can be obtained from the effective potential of LSM, such as,
\bea
\Omega_{LSM}(T,\mu)= \frac{-T\ln{\mathcal{Z}}}{V} = U(\sigma_l, \sigma_s)  + \Omega_{ \bar{q}q}(T, \mu_f),
\eea
Thus, the effective potential of the LSM is consisting of two parts:
\begin{itemize}
\item
the purely mesonic potential, which is given as a function of non-strange ($\sigma_l$) and strange  ($\sigma_s$) quark-flavor basis
 \begin{eqnarray}
U(\sigma_l, \sigma_s) &=& - h_l \sigma_l - h_s \sigma_s + \frac{m^2\, (\sigma^2_l+\sigma^2_s)}{2} - \frac{c\, \sigma^2_l \sigma_s}{2\sqrt{2}}  
+ \frac{\lambda_1\, \sigma^2_l \sigma^2_s}{2} +\frac{(2 \lambda_1
+\lambda_2)\sigma^4_l }{8}+ \frac{(\lambda_1+\lambda_2)\sigma^4_s}{4}. \hspace*{8mm} \label{Upotio}
\end{eqnarray}
\item
and the quark and antiquark contributions, which can be differentiated into two regimes:
\begin{itemize}
\item at zero magnetic field ($eB=0$) but finite temperature ($T$) and finite chemical potential ($\mu_f$) \cite{Kapusta:2006pm} 
\begin{eqnarray} 
\Omega_{ \bar{q}q}(T, \mu _f)&=& -2 \,T \sum_{f=l, s} \int_0^{\infty} \frac{d^3P}{(2 \pi)^3} \left\{ \ln \left[ 1+3\left(\phi+\phi^* e^{-\frac{E_f-\mu _f}{T}}\right) e^{-\frac{E_f-\mu _f}{T}}+e^{-3 \frac{E_f-\mu _f}{T}}\right] \right. \nonumber \\ 
&& \hspace*{35.5mm} \left.  +\ln \left[ 1+3\left(\phi^*+\phi e^{-\frac{E_f+\mu _f}{T}}\right) e^{-\frac{E_f+\mu _f}{T}}+e^{-3 \frac{E_f+\mu _f}{T}}\right] \right\}, \hspace*{8mm} \label{PloykovPLSM}
\end{eqnarray}
where $E=\sqrt{\vec{P}^2+m_f^2}$ is the dispersion relation of the valence quark and antiquark and $m_f$ is the $f$-th mass of quark flavor, which is respectively related to light and strange chiral condensates 
\bea
m_l=g \sigma_l/2, & & \qquad
m_s=g \sigma_s/\sqrt{2}, \label{mass_l_s}
\eea
where subscript $l$ refers to degenerate light quarks and $s$ to strange quark, and
\item
at nonzero magnetic field  ($eB\neq 0$) and finite $T$ and $\mu_f$,  the concepts of Landau quantization and magnetic catalysis, where the magnetic field is assumed to be oriented along $z$-direction, should be implemented.  
\end{itemize}
In doing of this, the approximation utilized constraints in the motion of particles along of the magnetic field with the transverse momentum of states $ \frac{|q_f|B}{2\pi}$ . One can apply one property of  magnetic catalysis \cite{Shovkovy:2013}, namely the dimensional reduction, $D \leftarrow D-2$ , so that integration over phase space becomes,
\bea
\int \frac{d^3P}{(2\pi)^3}  \longrightarrow   \frac{|q_f|B}{2\pi} \sum_\nu \int \frac{dP_z}{2\pi} (2-\delta_{0\nu}), \label{phaseeB}
\eea
where $\nu$ stands of the Landau quantization levels and $q_f$ is the corresponding electric charge of $f$-th quark flavors. Therefore, the quarks and antiquark potential at $eB\ne0$ becomes
\bea
\Omega_{\bar{q}q}(T, \mu _f, B) &=& - 2 \sum_{f=l, s} \frac{|q_f| B \, T}{(2 \pi)^2}\, \sum_{\nu = 0}^{\nu_{max_{f}}}  (2-\delta _{0 \nu})    \int_0^{\infty} dP_z \nonumber \\ && \hspace*{5mm} 
\left\{ \ln \left[ 1+3\left(\phi+\phi^* e^{-\frac{E_{B, f} -\mu _f}{T}}\right)\; e^{-\frac{E_{B, f} - \mu _f}{T}} + e^{-3 \frac{E_{B, f} -\mu _f}{T}}\right] \right. \nonumber \\ 
&& \hspace*{3.7mm} \left.+\ln \left[ 1+3\left(\phi^*+\phi e^{-\frac{E_{B, f} +\mu _f}{T}}\right)\; e^{-\frac{E_{B, f} +\mu _f}{T}}+e^{-3 \frac{E_{B, f} +\mu _f}{T}}\right] \right\}, \label{PloykovPLSMeB}
\eea
where $E_{B,f}$ is the dispersion relation of $f$-th quark-flavor,
\bea
E_{B, f} &=& \left[P_{z}^{2}+m_{f}^{2}+|q_{f}|(2n+1-\sigma) B\right]^{1/2}, \label{eq:moddisp}
\eea 
with $n$ is the quantization number known as the Landau quantum number and $\sigma$ is related to the spin quantum number, $\sigma=\pm S/2$. It is noteworthy  highlighting that, the quantity $2n+1-\sigma$ can be replaced by a summation over the Landau Levels $0\, \leq \nu\, \leq \nu_{max_f}$. The earlier is the Lowest Landau Level (LLL), while the latter stands for the Maximum Landau Level (MLL) $\nu_{max_{f}}$. For the sake of completeness, we mention that $2-\delta_{0 \nu}$ represents degenerate Landau Levels. $\nu_{max_{f}}$ contributes to the maximum quantization number ($\nu_{max_{f}} \rightarrow \infty$). The quark charges, the magnetic fields, the temperatures, and the baryon chemical potentials influence the maximum occupation of the Landau levels \cite{Boomsma:2010}.  In Ref. \cite{THD:magnetic}, we have discussed on how the Landau levels are filled up. Also, we have concluded that increasing Landau levels very slightly sharpens the phase transition and decreases the pseudo-critical temperature \cite{THD:magnetic}. The latter manifests inverse magnetic catalysis. 

The mechanism of the magnetic catalysis relies on a competition between valance and sea quarks \cite{Endrodi:2013, Fraga:2014pte, Andersen2014, Andersen2015, Bruckmann2013}. In light of this, when the magnetic field is switched on, the interaction of sea quarks leads to a net inverse magnetic catalysis. The magnetic field influences the quark condensates (sea quarks) and orders the Polyakov loops that imply suppression of the quark condensates (sea quarks) \cite{Endrodi:2013, Fraga:2014pte, Andersen2014, Andersen2015, Bruckmann2013}.   Applying finite magnetic field on QCD matter allows the description of important phenomenon such as CME and  magnetic catalysis  \cite{Ferreira2014q, Fraga:2014pte, Zhuang:2014syr, Ferreira2014s, Ferreira:2014we, Farias:2016}.

The inclusion  of the fermion vacuum term causes second-order phase transition in the chiral limit and other significant effects on the phase structure, which is commonly referred to as no-sea approximation \cite{Redlich:2010d} and  conjectured to distort the critical behavior, especially at the chiral phase transition.  When the fermion vacuum term is included, the transition can be of first- or second-order depending on the choice of coupling constants and on the density (chemical potential).  In Ref. \cite{Redlich:2010d}, the authors discussed on the thermodynamical differences between NJL and quark-meson (QM) models or LSM. In this approximation, the adiabatic trajectories obtained from the QM exhibit a {\it kink} at the chiral crossover phase transition, while from the NJL model, it is {\it smooth} everywhere \cite{blind}. This effect underlies a first-order phase transition from the LSM model in the chiral limit, especially when the fermionic vacuum fluctuations are neglected \cite{Redlich:2010as}.  

Equations (\ref{PloykovPLSM}) and (\ref{PloykovPLSMeB}) at vanishing and finite magnetic field, respectively, give the fermionic contributions to the QCD medium. The formal removal of the ultraviolet divergences is fully achieved through the fermion vacuum term \cite{Menezes:2009a} in sharp noncovariant cut-off ($\Lambda$)  \cite{Menezes:2009a},
\bea
\Omega^{\mbox{vac}}_{q\bar{q}} &=&  2 N_c N_f \sum_f \int_\Lambda \frac{d^3P}{(2\pi)^3}  E_f = \frac{-N_c N_f}{8\pi^2} \sum_f \left( m_f^4 \ln {\left[\frac{\Lambda + \epsilon_\Lambda}{m_f} \right]} - \epsilon_\Lambda \Lambda \left[\Lambda^2 + \epsilon_\Lambda^2\right]\right), \hspace*{5mm}
\eea
where $\epsilon_{\Lambda}=(\Lambda^2 + m_f^2)^{1/2}$ and $m_f$ is the flavor mass, with subscript $f$ running over light ($l$) and strange ($s$) quark flavors, Eq. (\ref{mass_l_s}). In finite magnetic field, we find that the fermion vacuum term has a negligible effect on the PLSM results.
\end{itemize}

The PLSM parameters, $m^2$, $h_l$, $h_s$, $\lambda_1$, $\lambda_2$, and $c$, besides the coupling $g$, the two condensates $\sigma_l$ and $ \sigma_s$ and the two order parameters $\phi$ and $\phi^*$ should be determined. The first six parameters can be fixed from experiments. Table \ref{tab:1a} summarizes the values of  these parameters at sigma mass $m_\sigma=800~$MeV \cite{Schaefer:2008hk}.

At finite volume ($V$) and finite magnetic field strength ($eB$), the free energy density is defined as  \hbox{$f=\mathcal{F}/V =-T \cdot \log [\mathcal{Z}]$}. From Eqs. (\ref{Uloop}), (\ref{Upotio}),  (\ref{PloykovPLSM}) and (\ref{PloykovPLSMeB}), the PLSM free energy density reads \cite{Tawfik:Magnetic2}
\begin{eqnarray}
f  &=&  U(\sigma_l, \sigma_s) +\mathbf{\mathcal{U}}(\phi, \phi^*, T) + \Omega_{  \bar{q}q}(T, \mu _f, B)  + \delta_{0,eB} \, \Omega_{ \bar{q}q}(T, \mu _f), \label{potential}
\end{eqnarray} 
where the last two terms of the right hand site represent the quark-antiquark contributions at finite and vanishing magnetic field, respectively. $ \delta_{0,eB}$ switches between both terms. Practically, only one of them shall be taken into account, separately.

In order to determine $\sigma_l$, $ \sigma_s$, $\phi$ and $\phi^*$, the free energy density, Eq. (\ref{potential}), should be minimized with respect to $\sigma_l$, $ \sigma_s$, $\phi$ and $\phi^*$, respectively 
\begin{eqnarray}\label{cond1}
\left.\frac{\partial f}{\partial \sigma_l}\right|_{min}, \quad \left.\frac{\partial
f}{\partial \sigma_s}\right|_{min}, \quad \left.\frac{\partial f}{\partial
\phi}\right|_{min}, \quad \left.\frac{\partial f}{\partial \phi^*}\right|_{min}.
\end{eqnarray} 
Although, the PLSM free energy density,  Eq. (\ref{potential}), is a complex function at finite chemical potential ($\mu\ne0$), its minimization becomes void of meaning. 
In general, any phase transition is characterized by an order parameters which are identified here with the expectation values of the sigma fields ($\sigma_l$, $ \sigma_s$) and  Polyakov loop potentials ($\phi$ and $\phi^*$). The behavior of the chiral order parameters are determined by the corresponding equation of motion. It is obtained by minimizing the real part of Eq. (\ref{potential}) ($\mbox{Re}\;f$) at a saddle point. In-medium condensates, this enables to determine the chiral order parameter of  phase transition with the corresponding fields as functions of $T$, $\mu$ and $eB$.

\section{Results and discussion \label{results}}

\subsection{Quark-hadron phase structure \label{structure}}

\subsubsection{Chiral and deconfinement order-parameters \label{prder_parameter}}

It is well accepted in the science community that the chiral phase structure for light and strange quark flavors is completely identified with their chiral order parameters. As we utilize PLSM, we first present their  associated meson condensates $\langle\sigma_l\rangle$ and $\langle\sigma_s\rangle$ and they are related to the quark condensates $\langle\bar{q}q\rangle$. Interested readers are kindly advised to consult the well-known textbook \cite{Kapusta:2006pm} for more details. In SU($2$), the sigma and pion fields are associated with bilinear forms of the quark fields, 
\bea
\bar{\sigma} \sim  \bar{q}q,  \qquad \qquad \vec{\pi} \sim i  \bar{q} \gamma_5 \vec{\tau} q.
\eea
It is apparent that the dimensions do not match, as if the mesonic condensates are given in GeV units, the quark condensates shall be given in GeV$^3$ units. Therefore, there must be some dimensional coefficients missed. It was suggested that they would be group invariant $\sigma^2 + \pi^2 \sim (\bar{q}q )^2 - ( \bar{q} \gamma_5 \vec{\tau} q)^2$ \cite{Kapusta:2006pm} and no corrections to this group invariance appear at higher orders of the temperatures $\mathcal{O}(T^2)$ \cite{Kapusta:2006pm, Bochkarev1986, Eletsky1993}. Furthermore, it was proved that these coefficients are temperature-independent \cite{Metzger:1994ss, Pirner:1994Conf}. Their values could be obtained in accordance from  partially conserved axial-vector current (PCAC) relations \cite{Narison:1989ax}. Such interconnections are essential in order to compare quark condensates to $\sigma_l$ and $\sigma_s$
\bea
\langle\bar{q}q\rangle &=& \frac{(-\epsilon_0)}{\sqrt{3}(2m_q+m_s)} \left(\sqrt{2}\sigma_l+\sigma_s\right) + 
                                                \frac{(-\epsilon_8)}{2\sqrt{3}(m_q-m_s)} \left(\sigma_l-\sqrt{2}\sigma_s\right) , \label{eq:qqsigm1}\\ 
\langle\bar{s}s\rangle &=& \frac{(-\epsilon_0)}{\sqrt{3}(2m_q+m_s)} \left(\sqrt{2}\sigma_l+\sigma_s\right) -
                                                \frac{(-\epsilon_8)}{\sqrt{2}(m_q-m_s)} \left(\sigma_l-\sqrt{2}\sigma_s\right) \label{eq:sssigm1},
\eea
where $\epsilon_0=0.02656~$GeV$^3$ and $\epsilon_8=-0.03449~$GeV$^3$ \cite{Metzger:1994ss}. The numerical estimation for both sets of quantities, $\langle\bar{q}q\rangle$ and $\langle\bar{s}s\rangle$ and $\sigma_l$ and $\sigma_s$, respectively, shows that they are almost identical, especially within at temperatures ${\cal O}(T_{c|\chi})$. From Eq. (\ref{mass_l_s}), the light and strange quark masses ($m_l$ and $m_s$) can be replaced by light and strange chiral condensates, respectively.  Consecutively, the chiral condensates ($\sigma_l$ and $\sigma_s$) and deconfinement order-parameters ($\phi$ and $\phi^*$) can be estimated from global minimization of the free energy density, Eq. (\ref{cond1}). Hence, the solution of the gap equations Eq. (\ref{cond1}) estimates the behavior of the condensates as a function of temperature, chemical potentials and magnetic field. For more details about the calculations of  $\sigma_l$, $\sigma_s$, $\phi$ and $\phi^*$ from the PLSM, the readers are kindly advised to consult Refs. \cite{Tawfik:LSM, Tawfik:Masses, Tawfik:Magnetic, Tawfik:quasi}. In doing this, we use $\sigma_{l_{0}}=92.4~$MeV and $\sigma_{s_{0}}=94.5~$MeV. Furthermore, PLSM can be utilized in determining the physical masses of the degenerate light and the  strange quarks under the assumption of degenerate quark chemical potentials, $\mu_u=\mu_d=\mu_s$. 

\begin{figure}[htb]
\centering{
\includegraphics[width=5.5cm,angle=0]{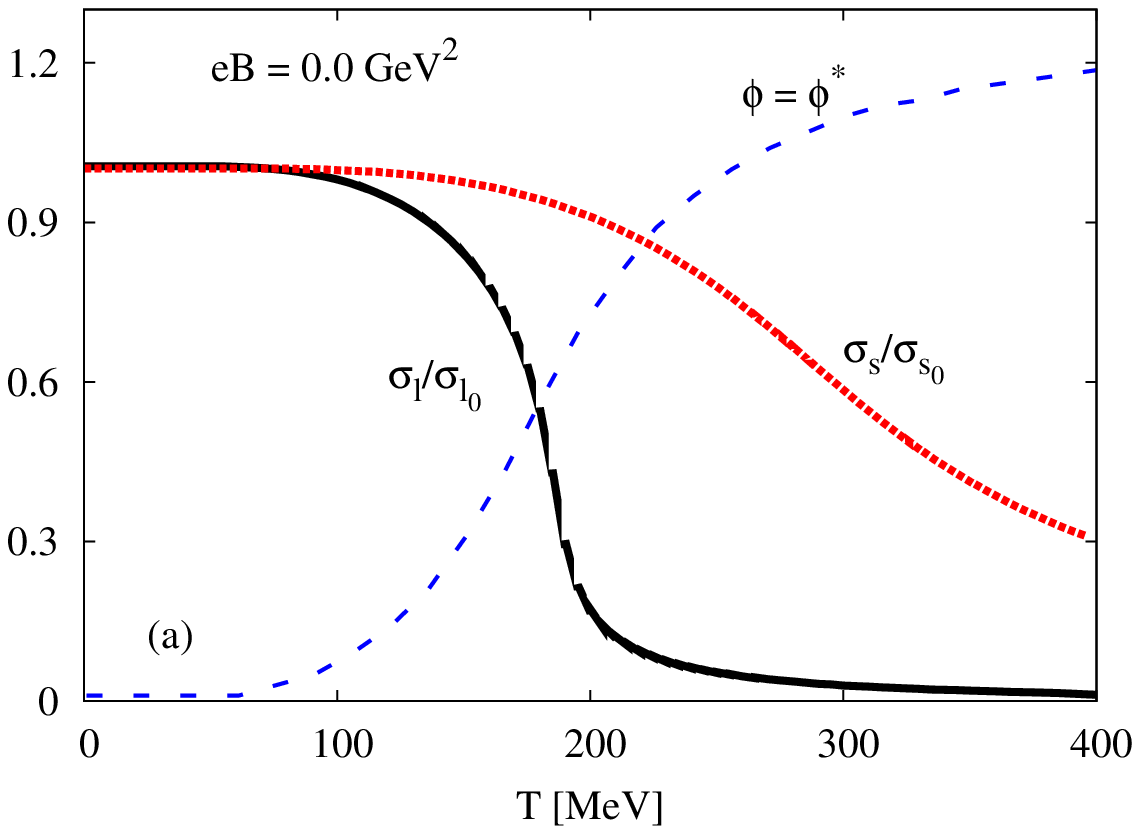}
\includegraphics[width=5.5cm,angle=0]{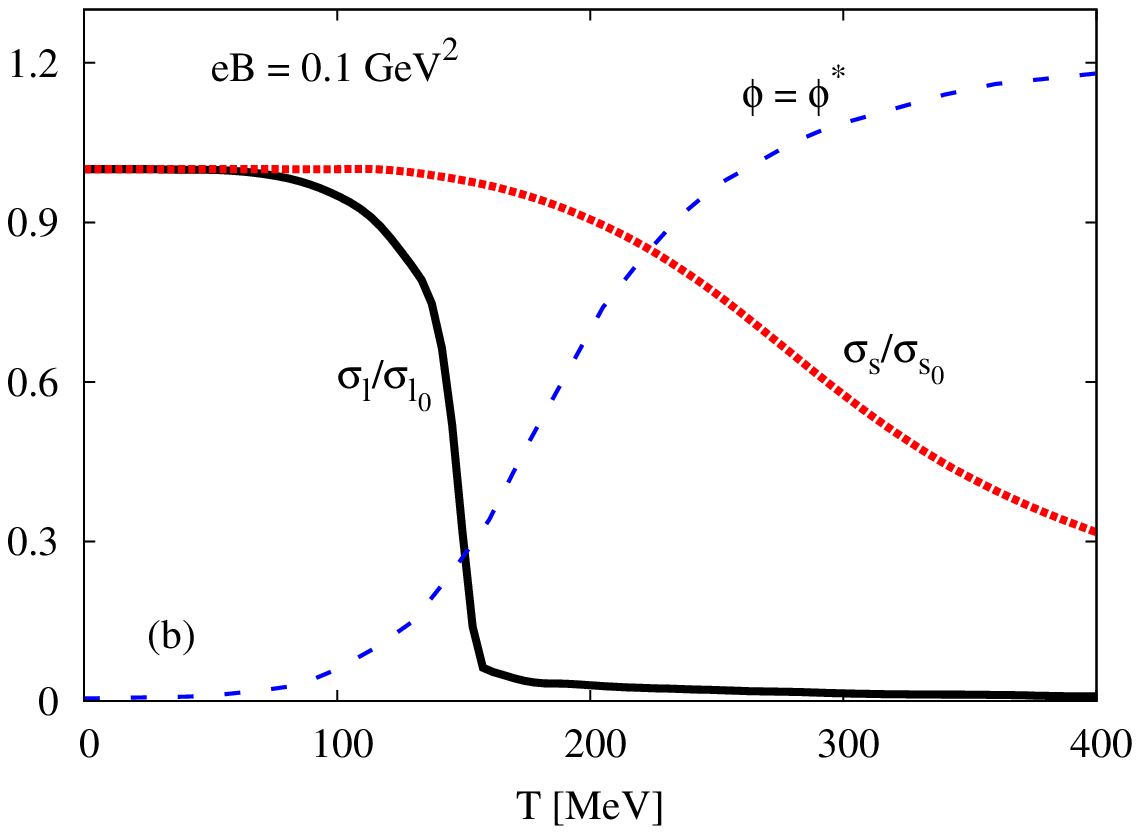}
\includegraphics[width=5.5cm,angle=0]{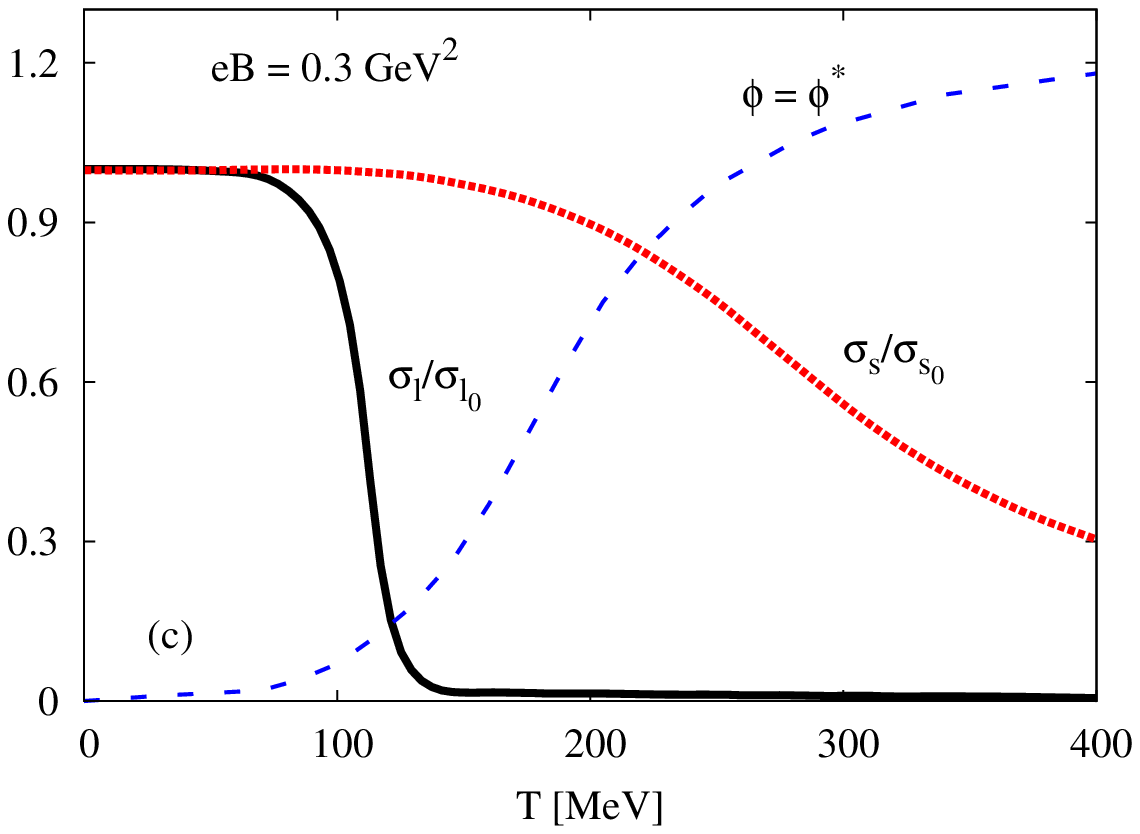}
\caption{\footnotesize The normalized chiral condensates $\sigma _l/\sigma _{lo}$ and $\sigma _s/\sigma _{so}$ (solid and dotted curves, respectively) and the Polyakov-loop potential, the order-parameters $\phi$ and $\phi ^*$  (dashed curve) are given as functions of temperature at different magnetic fields $eB=0.0$ [left-hand panel (a)], $eB=0.1$ [middle panel (b)] and $eB=0.3~$GeV$^2$ [right-hand panel (c)], respectively. \label{fig:cndst1} }
}
\end{figure}

Figure \ref{fig:cndst1} depicts the normalized chiral condensates, $\sigma_l/\sigma_{l_0}$ and  $\sigma_s/\sigma_{s_0}$, respectively, in presence of Polyakov-loop potentials, which - as mentioned - characterize the deconfinement phase-transition. At vanishing chemical potential, the two Polyakov-loop potentials are identical;  $\langle\phi\rangle=\langle\phi^{*}\rangle$. At $\mu=0$ but $e B\neq0$, both types of order parameters are estimated from the PLSM  in a wide range of temperatures. Fig. \ref{fig:cndst1} shows the temperature dependence of the chiral condensate and that of the deconfinement phase-transitions at  $eB=0.0$ [right-hand panel (a)], and at finite magnetic fields; $eB=0.1$ [middle panel (b)] and $eB=0.3~$GeV$^2$ [left-hand panel (c)], respectively. 

The pseudo-critical temperature ($T_\chi$) is defined as the temperature, at which the broken chiral-symmetry for light-quark condensate ($T_{\chi}^{(l)}$) and for strange-quark condensate ($T_{\chi}^{(s)}$) are restored. $T_\chi$  can be determined at the peaks of the temperature dependences of strange and nonstrange chiral condensates \cite{Tawfik:Magnetic}. 

Table \ref{table-chiral} summarizes the approximate estimation for $T_\chi$ at finite magnetic fields but $\mu=0$. It is conjectured that $T_{\chi}^{(l)}$ and $T_{\chi}^{(s)}$ denote the pseudo-critical temperatures at which the deconfinement phase-transition takes place. This observation was confirmed in lattice QCD simulations, especially at $\mu=0$.  As the magnetic field increases, we observe that $T_\chi$ of the light-quark condensate ($T_{\chi}^{(l)}$) decreases faster than that of the strange-quark condensate (moving from left-hand, to middle, to right-hand panel). In other words, increasing the magnetic field  contributes to the suppression in $T_\chi$, i.e. the restoration of the broken chiral symmetry takes place at lower temperatures. The earliness of the chiral phase transition relative to the lower temperatures (or the chiral condensate suppression) is known as the inverse magnetic catalysis (IMC).

\begin{table}[htb]
\begin{tabular}{|c|c|c|c|}
\hline 
 & $eB=0.0$ GeV$^2$ & $eB=0.1$ GeV$^2$ & $eB=0.3$ GeV$^2$ \\ 
\hline 
$T_{\chi}^{(l)} \;$[MeV]  & $177.8$&$150.4$ & $122.76$\\ 
\hline 
$T_{\chi}^{(s)}\;$[MeV]  & $218.6$ & $217.6$& $217.05$  \\ 
\hline 
\end{tabular} 
\caption{\footnotesize An approximate estimation for the chiral pseudo-critical temperatures at different magnetic fields. \label{table-chiral}  }
\end{table}

\subsubsection{Subtracted condensates}
\label{Sub_condensates}

The lattice QCD simulations combine both light- and strange-quark condensates \cite{lattice2008a}, 
\bea
\Delta_{l,s} &=&\frac{\left.\langle\bar{l}l\rangle - \left(\frac{m_l}{m_s}\right)\langle\bar{s}s\rangle\right|_T}{\left.\langle\bar{l}l\rangle - \left(\frac{m_l}{m_s}\right)\langle\bar{s}s\rangle \right|_{T=0} },
\label{subtracted1}
\eea
where $\langle\bar{q}q\rangle$ and $m_q$ stand for antiquark-quark condensate and quark mass, respectively, with $q\in[l,s]$. This quantity reads the ratio of net light- to strange-quark condensates at finite $T$ to the net of light- to strange-quark condensates at $T=0$ \cite{lattice2009b}. It is apparent that the dependence of the quark masses ($m_q$) on the corresponding chiral condensates at vanishing and finite temperature is essential.

In the present work, we introduce calculations for degenerate light-quarks and heavy strange-quark from SU($3$) PLSM. 
\bea
\Delta_{l,s} = \frac{\left. \sigma _l - \left(\frac{h_l}{h_s}\right) \sigma_s \right|_T }{\left. \sigma_l - \left(\frac{h_l}{h_s}\right) \sigma_s \right|_{T=0}},
\label{subtracted2}
\eea
where the flavor masses $m_l$ and $m_s$ are replaced by the explicit breaking strengths $h_l$ and $h_s$, receptively. The explicitly symmetry breaking terms $h_{i=l,s}$ can be obtained by Dashen-Gell-Mann-Oakes-Renner (DGMOR) relations \cite{DGMOR1, DGMOR2}, which relate the masses and decay widths of pion and kaon, $f_\pi$, $f_K$,  $m_\pi$, and $m_K$, respectively, such as \cite{Schaefer:2008hk}, 
\bea
h_l = f_\pi\, m_\pi^2,   \qquad  \qquad h_s = \sqrt{2} f_K m_K^2- \frac{f_\pi\,m_\pi^2}{\sqrt{2}}.
\eea 
However, we find that both are related to nonstrange and strange basis as $\bar{\sigma}_l=f_\pi$ and $\bar{\sigma}_s = (2f_K-f_\pi)/\sqrt{2}$, whereas $h_{i=l,s}$ are constants depending on the sigma mass ($m_\sigma$) \cite{Schaefer:2008hk}. They are listed out in Tab. \ref{tab:1a}.

Furthermore, the lattice QCD simulations define the chiral order-parameter in terms of the chiral condensate \cite{HotQCDtree},
\bea
M_b &=& \frac{m_s\, \langle\bar{q}q\rangle}{T^4}, \label{eq:Mb}
\eea
where $\langle\bar{q}q\rangle$ can be expressed in $\sigma_l$ and $\sigma_s$, Eq. (\ref{eq:qqsigm1}).

\begin{figure}[htb]
\centering{
\includegraphics[width=6.5cm,angle=0]{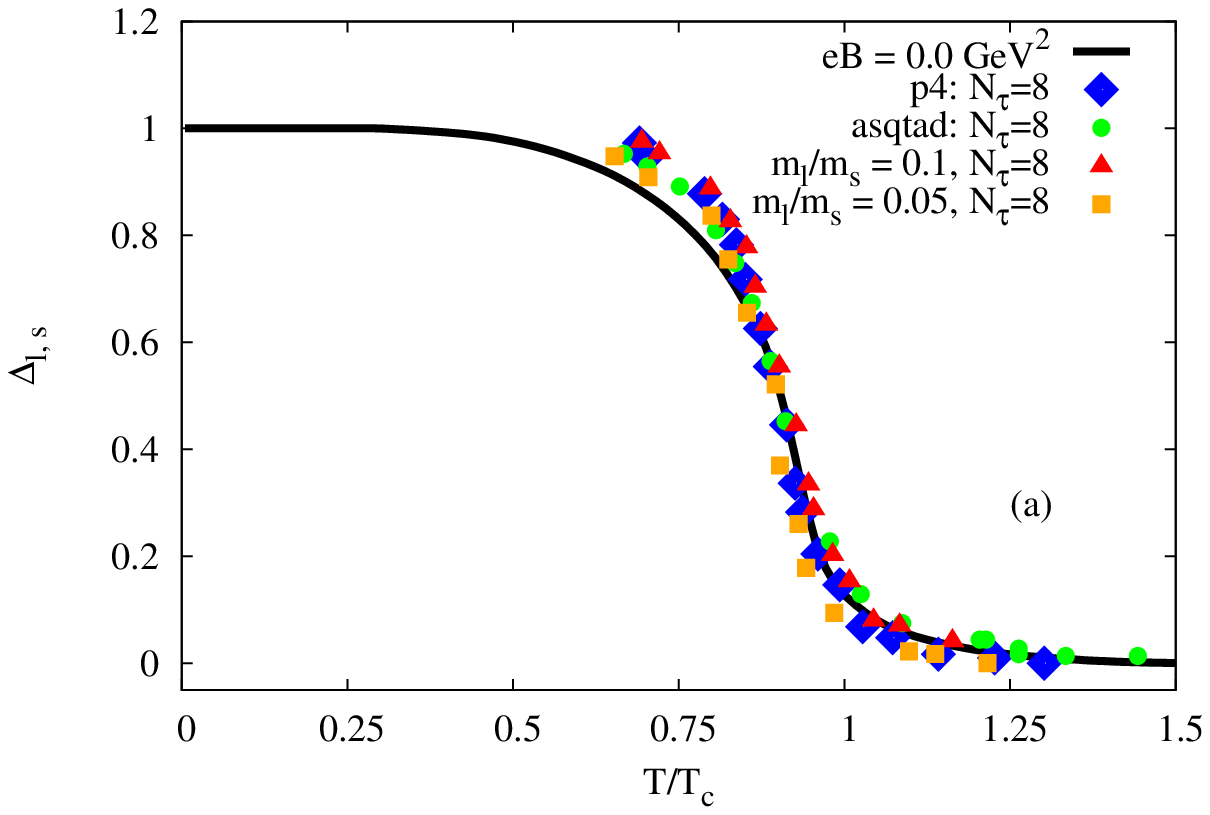}
\includegraphics[width=6.5cm,angle=0]{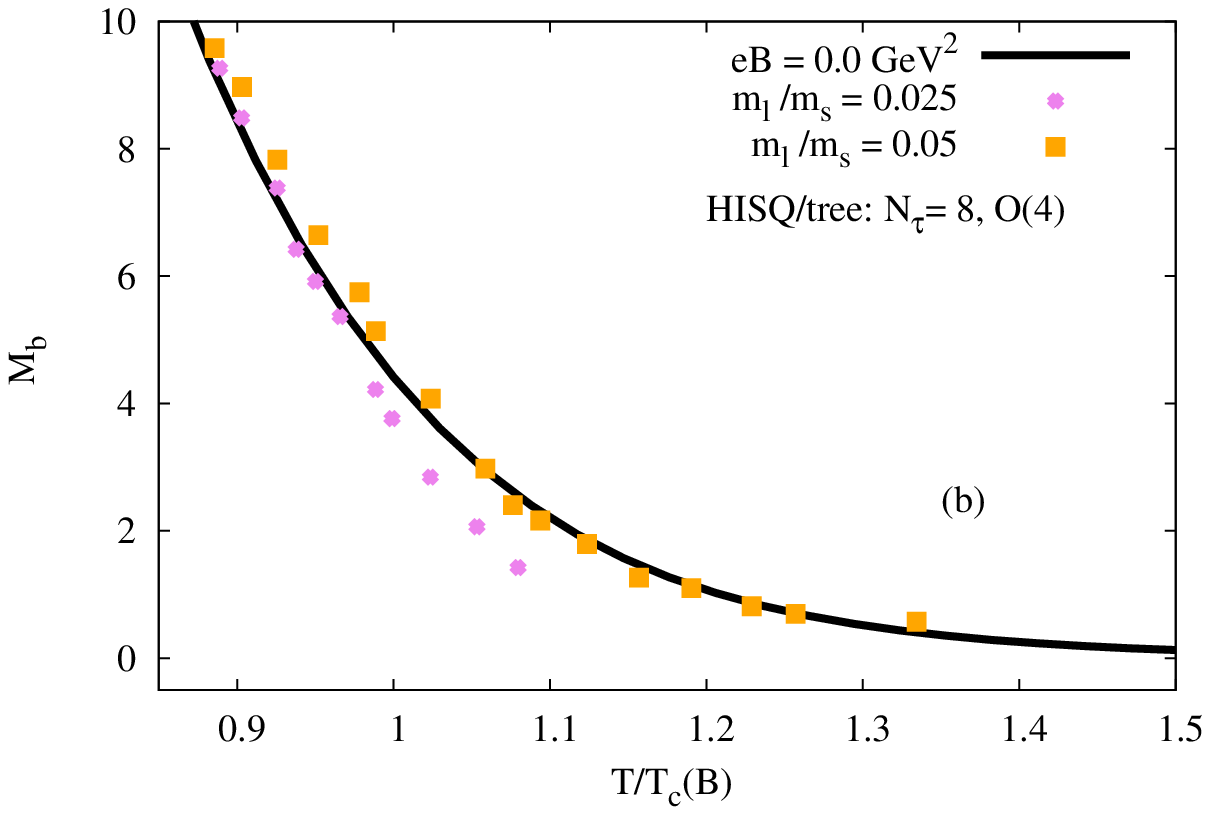}
\caption{\footnotesize Left-hand panel (a): the subtracted condensate (light- and strange-quark net-condensate) $\Delta_{l,s}$ given as a function of temperature at $\mu=0$ and $eB=0.0~$GeV$^2$ (solid curves) compared to lattice QCD calculations (symbols) \cite{lattice2009a,lattice2009b}. Right-hand panel (b): the temperature dependence of the chiral order-parameter, $M_b$, Eq. (\ref{eq:Mb}) is presented at  $eB=0.0$ (solid curves) compared with HISQ/tree lattice QCD with $N_{\tau}=8$ and two values for the quark masses, $M_q/M_s=0.025$ (close circle) and $M_q/M_s=0.05$ (close square) \cite{HotQCDtree}.
\label{fig:sbtrc1}}} 
\end{figure}

The left-hand panel of Fig. \ref{fig:sbtrc1} (a) presents the subtracted chiral-condensates as a function of temperature at vanishing chemical potential and magnetic field (solid curves) and compares the PLSM results with various ($2+1$) lattice QCD simulations, in which asqtad \cite{Orginos:2009} and p4 \cite{Heller:1999, Peikert:1998} improved staggered fermion actions with almost physical strange and light-quark masses and temporal extension $N_\tau=8$ are utilized. The agreement between both sets of calculations is excellent. The steeper decline in the subtracted chiral-condensate comes from the pure gauge potential in the grand canonical calculations in absence of the gluons interactions. We observe that $\Delta_{l,s}$ possesses a constant finite value at low $T$. When the temperature approaches about half $T_c$, $\Delta_{l,s}$ begins to decrease with increasing temperature. This signal indicates the liberation of the quark and gluon degrees-of-freedom. The hadron-quark deconfinement and the restoration of the broken chiral symmetry are conjectured to take place at $T_c$.

It is noteworthy emphasizing that the introduction of the Polyakov-loop correction improves the calculations of the pure gauge potential through improving the gluon contributions to the interactions. The latter leads to a smooth chiral transition or a repaid crossover \cite{Orginos:2009, Heller:1999, Peikert:1998}. 

The right-hand panel of Fig. \ref{fig:sbtrc1} (b) shows the temperature dependence of $M_b$. As introduced, $M_b$, Eq. (\ref{eq:Mb}), combines the mass of the strange quark ($m_s$) with the light-quark condensate ($\sigma_l$) normalized to $T^4$. The PLSM calculations are compared with HISQ/tree lattice QCD with $N_\tau=8$ and two values for the quark masses, $M_q/M_s=0.025$ (close circle) and $M_q/M_s=0.05$ (close square) \cite{HotQCDtree}. The agreement between the two sets of calculations is convincing, especially at vanishing magnetic field (solid curve) and in the crossover region, in which the rapid decline in $M_b$ indicates a smooth phase transition from finite to vanishing $M_b$.

\subsection{Thermodynamic quantities \label{thermo}} 

Various thermodynamic quantities can be deduced from  the total free energy density, Eq. (\ref{potential}). For instance, the pressure at vanishing $\mu$ but finite $T$ and $e B$ defines the absolute free energy density of the system of interest
\bea
p(T, eB)=-f(T, e B).
\eea  
At $eB=0$ and $\mu=0$, the free-energy density $f(T, e B)=\mathcal{F}(T, e B)/V$ can be given as 
\bea
f(T, e B)=\epsilon(T, e B) -T\, s(T, e B),
\eea
where $\epsilon$ and $s$ being energy density and entropy density, respectively. At finite magnetic field ${\bf B}=B\, \hat{e}_z$  directed along the $z$-axis, the free energy density can be given as \cite{Landau:1995}
\bea
\mathfrak{f}&=& \epsilon^{\mbox{tot}} - \epsilon^{\mbox{field}} - T s = \epsilon^{\mbox{tot}} - T s - e B \; \mathcal{M}, \label{totalfree}
\eea
where $\epsilon^{\mbox{tot}}=\epsilon + \epsilon^{\mbox{field}}$  is the total energy density consisting of the energy density ($\epsilon$) characterizing the system of interest and $\epsilon^{\mbox{field}}=e B \mathcal{M}$ stemming from the influence of the magnetic field, with $\mathcal{M}$ being the magnetization and the magnetic field $e B$ is given in units of the elementary charge $|e|>0$. 

We recall that at vanishing magnetic field, the partition function can be given by an integral over the six-dimensional phase space and an integral over the dispersion relations. In this case, the energy-momentum dispersion relation, $E_f$, follows the Lorentz invariance principle, but at finite magnetic field, the integral is dimensionally reduced and simultaneously accompanied by a considerable modification in the dispersion relations, themselves. At $eB\neq0$, the velocity of a test particle with the momentum $\partial  \epsilon^{\mbox{tot}} /\partial p$ can be deduced from the dispersion relations,
\bea
v_{P_z} &=&  c \left[\frac{c\, P_z}{c\, P_z + 2 |q_f|(\nu+\frac{1}{2}-\frac{\sigma}{2}) B}\right].
\eea
The Lorentz invariance principle can be confirmed through the velocity of a test particle with the momentum $P$. The causality is apparently guaranteed as $v_{P_z}$ doesn't exceed the speed of light ($c$), i.e. as long as the $eB$-term is finite positive, which should be estimated, quantitatively, as a function of temperature and magnetic field.

Form Eq. (\ref{totalfree}), the entropy density and the magnetization can be obtained as 
\bea
s &=& - \frac{1}{V} \frac{\partial f}{\partial T}, \label{eq:entropy} \\
\mathcal{M} &=& - \frac{1}{V} \frac{\partial f}{\partial (eB)}. \label{eq:Magnet}
\eea 
As the magnetic field marks a preferred direction, the pressure $p_i$ might be different along the geometrical effect of the magnetic field. The magnetic field is directed along the $z$-direction. The volume of the system reads $V= L_x  \cdot L_y \cdot L_z$.  Thus, lattice QCD simulations distinguish between two different systems: 
\begin{itemize}
\item B-scheme: the magnetic field is kept fixed in all direction leading to an isotropic pressure 
\bea
p_x = p_y = p_z.
\eea
\item $\Phi$-scheme: the magnetic flux ($\Phi=e B \cdot L_x\, L_y$) is kept fixed leading to an anisotropic pressure 
\bea
p_x = p_y = p_z - eB \cdot \mathcal{M}.
\eea 
\end{itemize}
Accordingly, the thermodynamic quantities should be corrected. For instance, the trace anomaly (interaction measure) in $\Phi$-scheme becomes
\bea
I &=& \epsilon - 3p_z + 2  eB \cdot \mathcal{M}. \label{InteractionMeasure}
\eea  

For the hydrodynamical investigations of relativistic heavy ion collisions, the speed of sound squared ($c_s^2$) is an essential thermodynamic quantity that can be related with the equation of state,  $p(\rho)$  \cite{lattice2009a}.  
\bea
c_s^2 &=& \left.\frac{\partial p}{\partial \epsilon}\right|_{s/n}  = \epsilon \frac{\partial}{\partial \epsilon} \left( \frac{p}{\epsilon} \right)+ \left( \frac{p}{\epsilon} \right). \label{eq100:cs2}
\eea
The speed of sound squared at fixed entropy per particle, $s/n$, i.e. adiabatic $c_s^2$, is given by  
\bea
c_s^2 &=& \frac{\partial p}{\partial \epsilon}\Bigg\vert_{s/n} \equiv 
\frac{\frac{\partial p}{\partial T}\Bigg\vert_V}{\frac{\partial \epsilon} {\partial T} \Bigg\vert_V}
=\frac{s/T^3}{c_V/T^3}, \label{eq:cs2}
\eea
where $s=S/V$ is  the entropy density and $n=N/V$ is the particle density. With this regard, it should be noticed that there is a very fine difference between this adiabatic $c_s^2$ (at fixed $s/n$) and the isentropic $c_s^2$, which is characterized by fixed $s$. Accordingly, one can depending on the thermodynamic quantities that are kept fixed determine {\it different} speeds of sound squared \cite{Piattella:2013wpa}. The equivalence of $c_s^2$ with $s/c_V$ is only at $\mu=0$, where the specific heat, $c_V=(\partial \epsilon/\partial T)|_V$, determines the thermal variation of the energy density at constant volume $V$. In the present work, we are calculating adiabatic $c_s^2$. 

At $\mu\neq0$ and fixed $s/n$, the speed of sound squared reads \cite{Piattella:2013wpa}
\begin{eqnarray}
c_s^2 &=& \frac{n}{p+\epsilon}\frac{n\, \chi_{T T} - 2\, s\, \chi_{\mu T} + \frac{s^2}{n}\, \chi_{\mu \mu}}{\chi_{T T}\chi_{\mu \mu}-(\chi_{\mu T})^2},
\end{eqnarray}
where $\chi$ is the second derivative, known as susceptibility, with respect to the given subscripts. In physics,   the susceptibility - depending on whether same or different quantum numbers are taken into consideration, such as strangeness and baryon quantum number - represents fluctuations or correlations, respectively. The present calculations are devoted to fluctuations, as only the baryon quantum numbers are taken into account. It is obvious that $\chi_{\mu T}\equiv \partial^2 p/\partial \mu\partial T$ is closely related to $c_V$, while $\chi_{\mu \mu}\equiv \partial^2 p/\partial \mu\partial \mu$ and $\chi_{TT}\equiv \partial^2 p/\partial T\partial T$ to the particle number susceptibility and the thermal variation of the entropy, respectively.  From this extension, we gain another confirmation for the conclusions drawn in Ref. \cite{Tawfik:2012ty} that the fluctuations come up with remarkable contributions to $c_s^2$ at $\mu\neq0$.

The Stefan-Boltzmann (SB) limits can be deduced from the lowest-order perturbation theory. For massless quarks and gluons \cite{Kapusta:2006pm}
\begin{eqnarray}
T\, \log \, \mathcal{Z}(V,T,\mu) &=& \frac{g_f\, V}{12} \left(\frac{7}{30}\pi^2\, T^4 + T^2\, \mu^2 +\frac{1}{2\, \pi^2} \mu^4 \right) + g_b\, V \frac{\pi^2}{90} T^4, \label{eq:Boltzm1}
\end{eqnarray}
where $g_f$ and $g_b$ are the degrees of freedom of quarks and gluons, respectively. At finite magnetic field ($e B\neq0$) \cite{LQCD:Magnet2014}
\bea
T\, \log \, \mathcal{Z}(V,T,\mu, eB) = \frac{19\, V \pi^2}{36} T^4 + b_1^{\mbox{free}} (eB)^2\, V \, \log\left(\frac{T}{\Lambda_H}\right)+\cdots,
\eea
where $b_1^{\mbox{free}}$ is a parameter to be fixed by leading-order perturbation theory and $\Lambda_H=0.12~$GeV is a renormalization scale \cite{LQCD:Magnet2014}, which exclusively appears in the perturbation terms.

The present work is devoted to analyzing the quark-hadron phase structure, the thermodynamics and the magnetisation of the QCD matter at finite temperatures. To conduct this ambitious project, we have to show first that our QCD-like model, the PLSM, is  capable to reproduce recent first-principle lattice QCD calculations. This includes the determination of various PLSM parameters as introduced in earlier sections.

It is in order now to elaborate the main characterizations of the lattice QCD simulations utilized in the present work. 
\begin{itemize}
\item At finite magnetic field, various ($2+1$) lattice QCD thermodynamic quantities with physical quark masses have been determined \cite{LQCD:Magnet2014}, as given in right-hand panel of Figs. \ref{trace}, \ref{fig:cs2}, \ref{fig:scv} and Figs.  \ref{fig:Magn_TceBA} and  \ref{fig:Magn_TceBC}. These are sensitive to the change in the control parameters of the system, which include temperatures, chemical potentials corresponding to the various conserved charges and, as in the present work, the finite magnetic field along the longitudinal direction ${\bf B}=B\; \hat{e}_z$. The lattice calculations use tree-level improved Symanzik gauge action, and stout improved staggered quarks in the fermionic sector, fixed ratio of light- and strange-quark masses, $m_l=m_s/28$ and lattice size $N_{\sigma}^3 \times N_{\tau}$.

\item The lattice QCD results presented in the left-hand panel of Figs. \ref{trace}, \ref{fig:cs2}, \ref{fig:scv} and \ref{fig:cv} (open circle) are  HotQCD simulations  with $2+1$ flavor at finite temperature \cite{lQCD2014}. They are reliably extrapolated to the continuum limit. Various thermodynamic quantities are deduced as  functions of temperature at vanishing chemical potential. Their reliable continuum extrapolated results on the equation of state have been performed with stout-smeared staggered action \cite{action01}. At high temperature, this action effectively reduces taste symmetry violation effects, which remain large for ${\cal O}(a^2)$ cutoff, where $a$ is the lattice spacing \cite{action01}. The current continuum extrapolated results \cite{lQCD2014} are obtained with the HISQ action. They show an excellent agreement with the stout action \cite{action01}. The light-quark mass is fixed as a fraction of the strange-quark mass $\hat{m}_l =\hat{m}_s/20$. The results are obtained at lattice size $N_\sigma^3 \cdot N_{\tau}$, where $N_{\tau}$ is temporal extent and $N_\sigma$ is the spatial dimension.

\item The ($2+1$) lattice QCD calculations are obtained at finite temperature with physical strange-quark mass and almost physical light-quark masses $\hat{m}_l=\hat{m}_s/10$ \cite{lattice2009a}. Two different improved staggered fermion actions; the asqtad and p$4$ actions, are implemented with temporal and spacial extension $N_{\tau}=8$ and $N_{\sigma}=32$, respectively.

\end{itemize}

It is apparent that the lattice QCD simulations at $eB\ne0$  \cite{LQCD:Magnet2014} give smaller quantities than the ones estimated at $e B=0$ \cite{lQCD2014}. At $eB\ne0$, the system undergoes modifications due the presence of finite magnetization in the region of  crossover region, where the magnetic field contribution to the suppression in the chiral condensate become large. To express the free energy as an integral over the longitudinal momentum, which is directed towards $z$-axis, i.e. ${\bf B}=B\hat{e}_z$, and a summation over the {\it quantized} Landau levels, both should be performed numerically corresponding to Eqs. (\ref{PloykovPLSMeB}) and (\ref{potential}).  Due to the occupation of the Landau levels, the thermal free-energy can be suppressed as the magnetic field grows \cite{THD:magnetic}.  

Some thermodynamic quantities shall be elaborated in the sections that follow. This includes the trace anomaly, the speed of sound squared, the entropy density , and the specific heat.

\subsubsection{Trace anomaly}

The normalized trace anomaly, which can directly be derived from the trace of the energy-momentum tensor $T_{\nu}^{\mu}$, at vanishing magnetic field, the trace anomaly   $I= \epsilon - 3 p$, where $\epsilon\, (p)$ being energy density (pressure), determines the degrees of freedom and helps in deducing further thermodynamic quantities such as energy density, pressure, and entropy density of the system of interest. In QCD, the trance anomaly can be related to the strong coupling constant, $\propto T^4\, \alpha_s^2$ \cite{Tawfik:2013eua}. This quantity vanishes in scale invariant theory such as massless, collisionfree gas of quarks and gluons. For a correlated gas (with nonvanishing interactions), the QCD scale parameter remains finite. The QCD asymptotic freedom implies that the strength of the interaction weakens with increasing temperature [far above the pseudo-critical temperature ($T_c$)]. At temperatures smaller than $T_c$, the trace anomaly increases with the increase in the temperature because more massive hadrons become relevant. 

\begin{figure}[htb]
\centering{
\includegraphics[width=5.5cm,angle=-90]{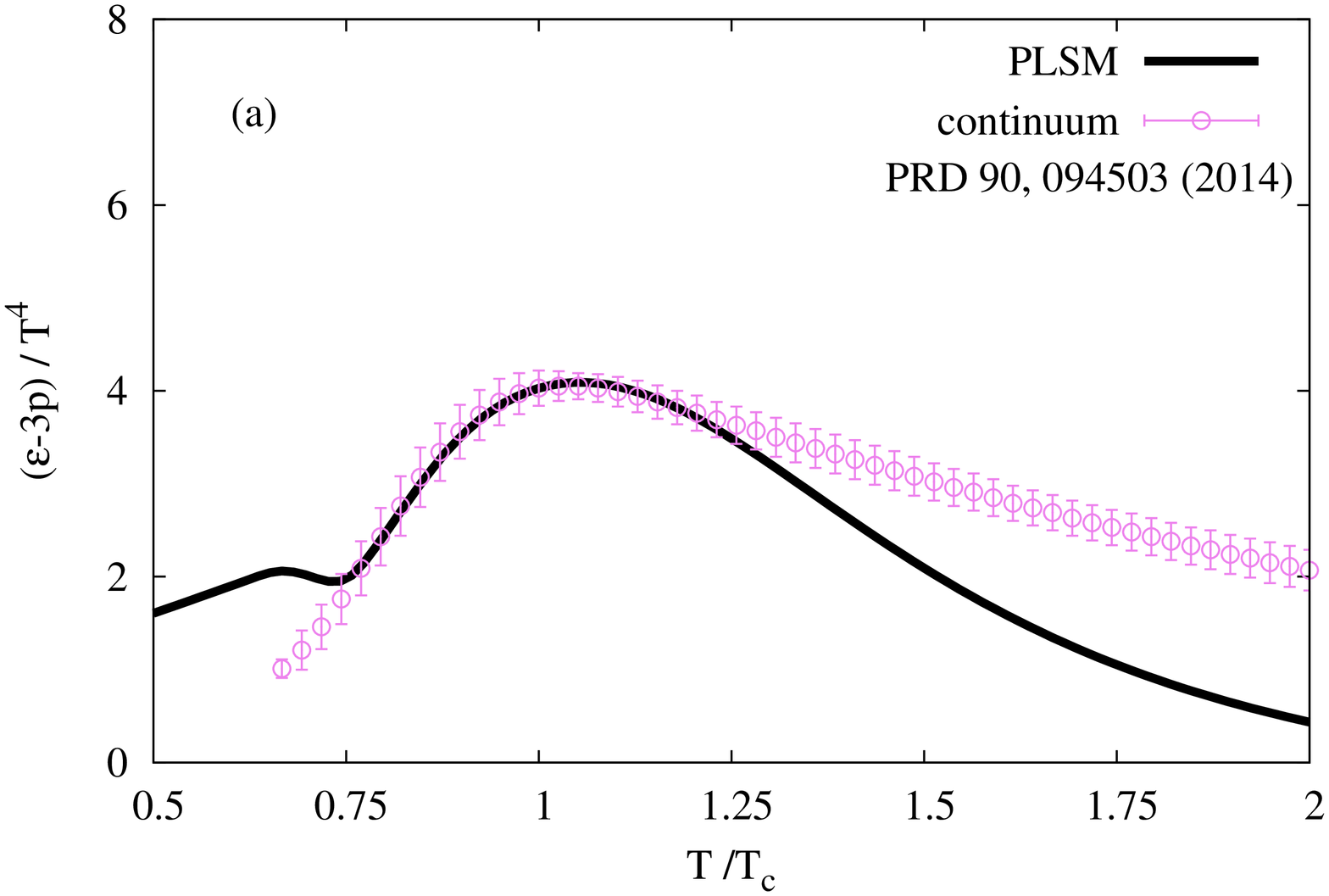}
\includegraphics[width=5.5cm,angle=-90]{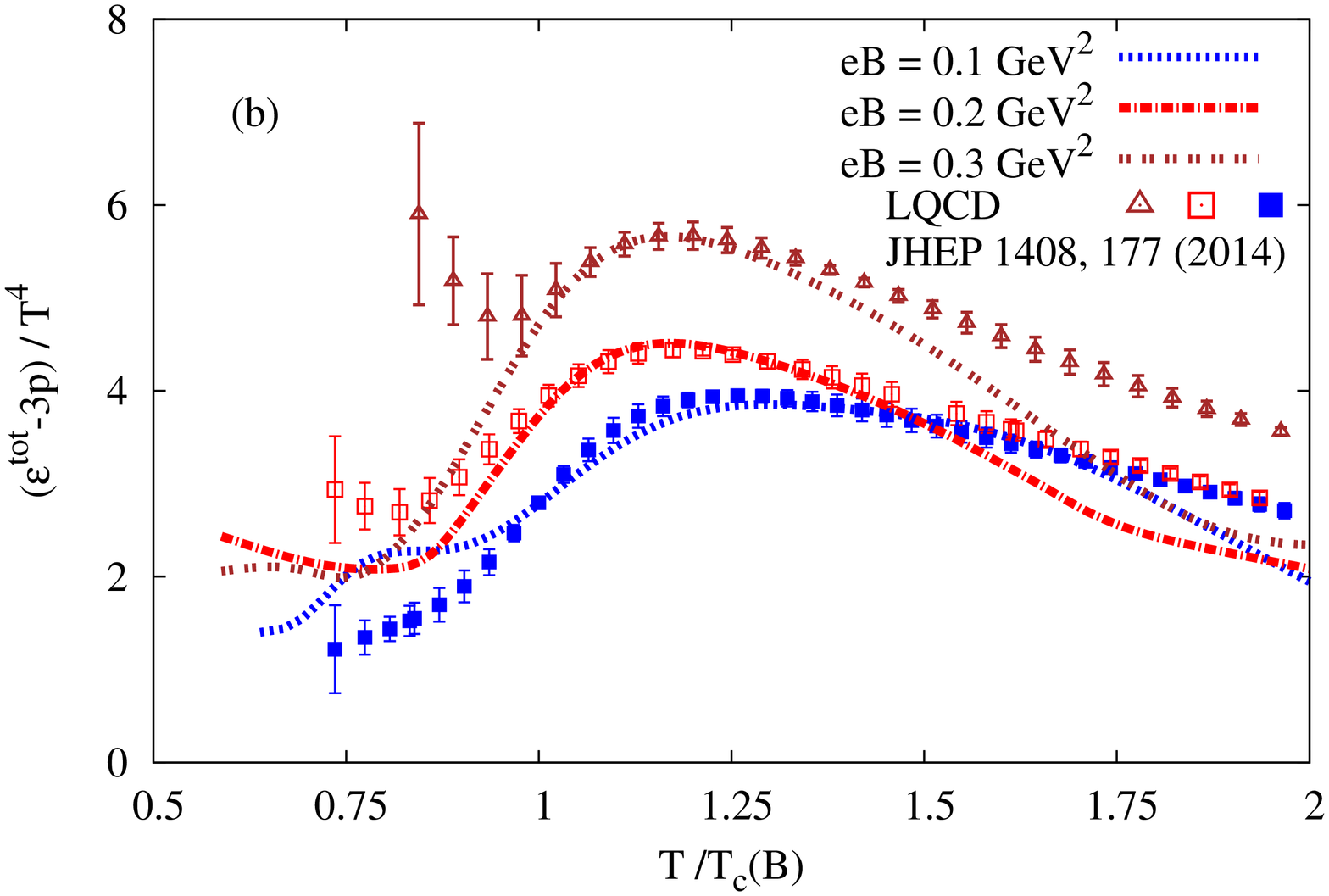}
\caption{\footnotesize Left-hand panel (a): the normalized trace-anomaly, $(\epsilon - 3p)/T^4$, as a function of temperature compared with lattice QCD calculations extrapolated to the continuum limit are given as open circles \cite{lQCD2014}.  Right-hand panel (b): the normalized trace anomaly at different values of magnetic fields $eB=0.1$ (dotted curve), $eB=0.2$ (dotted-dash curve)  and $eB=0.3~$GeV$^2$ (double-dotted curve) compared with the recent lattice QCD (close square), (open square) and (open triangle), respectively \cite{LQCD:Magnet2014}.  
\label{trace}}}
\end{figure}%

In Fig. \ref{trace}, the temperature dependence of the normalized trace-anomaly calculated from PLSM at vanishing [left-hand panel (a)] and nonzero magnetic field [right-hand panel (b)]. In left-hand panel (a), the trace anomaly normalized to $T^4$ is given in comparison with lattice QCD calculations extrapolated to continuum  limit \cite{lQCD2014}. There is a well agreement with the lattice QCD simulations, especially in the hadronic phase. This is also obvious near the phase transition region. At higher temperatures, there is a repaid decrease in the interaction. This seems to be faster than the one in the lattice QCD simulations. 

In right-hand panel of Fig. \ref{trace} (b), the modified normalized trace anomaly, Eq. (\ref{InteractionMeasure}), is depicted as a function of temperature. The thermodynamic modifications due to the presence of finite magnetic field have been discussed in the previous section. Our PLSM calculations are also given at different values of magnetic fields; $eB=0.1$ (dotted curve), $eB=0.2$ (dotted-dash curve)  and $eB=0.3~$GeV$^2$ (double-dotted curve)  and compared with recent lattice QCD \cite{LQCD:Magnet2014} (close square),  (open square) and (open triangle), respectively. 

As discussed in the previous section, the thermodynamic quantities, which are derived from the free energy, are accordingly modified, especially due to the presence of finite magnetization. The latter measures the response of the system of interest to applying finite magnetic field, i.e. magnetization. At $e B\neq0$, the free energy from PLSM is calculated as an integral over the longitudinal momentum along the direction of magnetic field ${\bf B}=B\hat{e}_z$ and a summation over the {\it quantized} Landau levels. Adding finite magnetization for the free energy affects various thermodynamic quantities. Furthermore, due to the occupation of the Landau levels, the thermal free energy suffers from an additional suppression with increasing magnetic field. Thus, the total energy-density likely differs from the quantity depicted in the left-hand panel (a). 

In the hadron phase, the trace anomaly becomes very small at low temperature, but increases with the increase in the temperature. A peak appears at the pseudo-critical temperature. A further increase in the temperature decreases the trace anomaly, i.e. derives the system strongly into the deconfined status. The trace anomaly signals breaking of the scale invariance in the system of interest. Its peak around $T_c$ is conjectured to signal anomalous thermodynamic properties, for example, the bulk viscosity, which measures how easy or difficult for the system to relax back to equilibrium after going through scale transformation. This would mean that the peak of the trace anomaly refers to maximum bulk viscosity, that is likely when the breaking of scale invariance becomes maximum.

\subsubsection{Speed of sound squared}

\begin{figure}[htb]
\centering{
\includegraphics[width=5.5cm,angle=-90]{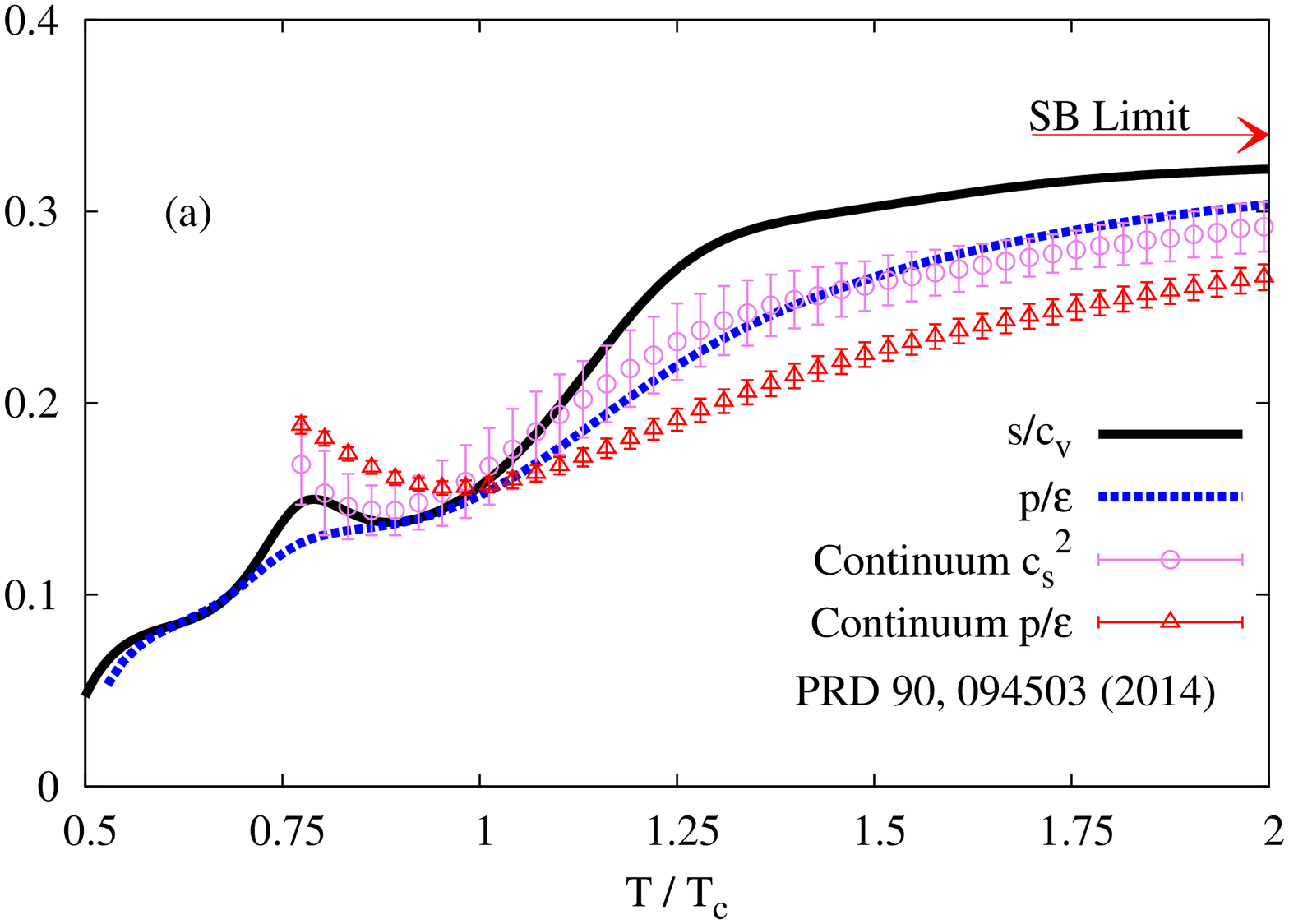}
\includegraphics[width=5.5cm,angle=-90]{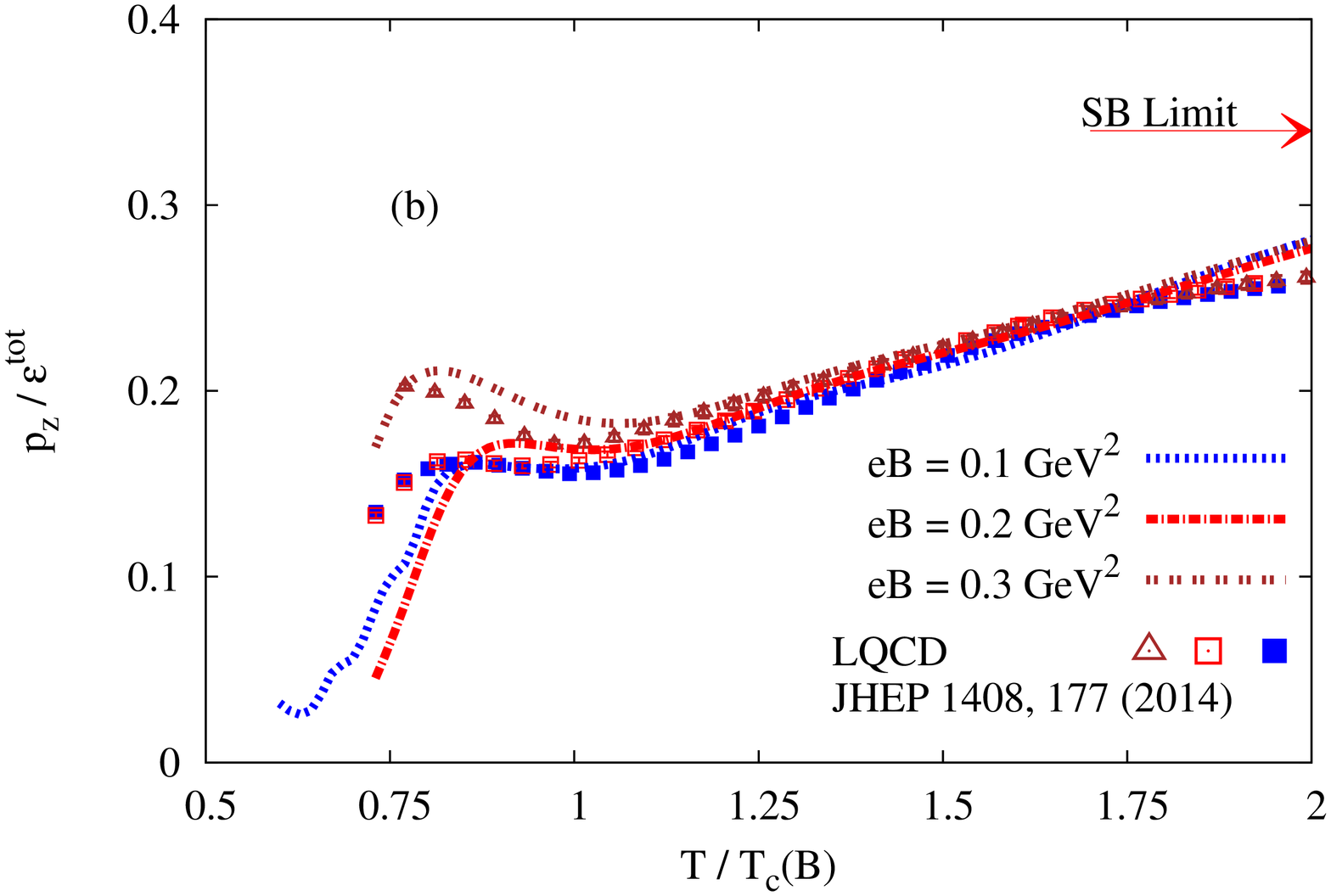}
\caption{\footnotesize Left-hand panel (a): the speed of sound squared is calculated as a function of temperature from $s/c_v$ (solid curve) and $p/\epsilon$  (dotted curve) compared with lattice QCD calculations extrapolated to the continuum limit presented as open circles \cite{lQCD2014}. Right-hand panel (b): at finite magnetic field, the ratio $p/\epsilon$ at $eB=0.1$ (dotted curve), $eB=0.2$ (dotted-dash curve)  and $eB=0.3~$GeV$^2$ (double-dotted curve) is compared with the recent lattice QCD (close square), (open square) and (open triangle), respectively \cite{LQCD:Magnet2014}.  
\label{fig:cs2}}}
\end{figure}

Analogy to the hydrodynamical approaches, which have been applied on the relativistic heavy-ion collisions and led to the RHIC discovery of the new-state-of-QCD-matter \cite{mgRHIC}, the speed of sound squared ($c_s^2$) plays an essential role in estimating the equation of state $p(\rho)$. As discussed in Ref. \cite{Tawfik:2012ty}, the definition of $c_s^2$ as in Eq. (\ref{eq:cs2}) means that the energy fluctuations and other collective phenomena associated with the specific heat are not taken into account.  

The speed of sound squared is related to the trace anomaly ($I=\epsilon-3p$), where the conformal measure is given as
\bea
\mathcal{C} &=& \frac{I}{\epsilon} = \frac{\epsilon -3p}{\epsilon} \approx 1 - 3 c_s^2,
\eea 
which becomes maximum around the region of the phase transition, in which the speed of sound squared becomes small. At high temperatures, the speed of sound squared reaches the Stefan-Boltzmann (SB) limit of $1/3$. In this limit, the conformal measure obviously vanishes. In left-hand panel of Fig. \ref{fig:cs2} (a), the temperature dependence of the speed of sound squared calculated from the ratio $s/c_v$ and $p/\epsilon$ are compared with recent lattice QCD calculations with continuum extrapolation  \cite{lQCD2014}. The right-hand panel (b) illustrates the ratio $p/\epsilon$ as a function of temperature and compares the PLSM calculations with the recent lattice QCD in finite magnetic field \cite{LQCD:Magnet2014}. It is remark-worthy  highlighting that our calculations are based on $\partial p/\partial \epsilon$, but for a better scripting we abbreviate this to $p/\epsilon$. 

Left-hand panel of Fig. \ref{fig:cs2} (a) shows the PLSM results on $p/\epsilon$ (dashed curve) and on $s/c_v$ (solid curve) and compares them with lattice QCD calculations extrapolated to the continuum limits  \cite{lQCD2014}. The lattice QCD results on both $c_s^2$ (open circles) and $p/\epsilon$(open tingles) are extrapolated to the continuum limits.  

Both quantities $p/\epsilon$ and $s/c_v$ are apparently confirmed from in the continuum extrapolation of the lattice QCD  \cite{lQCD2014}. Above $T_c$, the PLSM calculations seem to overestimate $s/c_v$. At lower temperature, an opposite temperature-dependence is observed. It is apparent that $c_s^2$ (solid curve) matches $p/\epsilon$ (dashed curve). The PLSM and lattice QCD calculations confirm that there is a difference between  $c_s^2$ (open circles) $p/\epsilon$ (open tingles). While an agreement is improved in the hadron phase, this becomes poor in high temperature. In the high-temperature region, the PLSM overestimates the lattice calculations.
  
It is noteworthy highlighting that the temperature dependence of $c_s^2$ seems to be sensitive to the normalization to $T_c$. The lattice calculations are given in dependence on the temperature ($T$), which is then normalized to $T_c \simeq 181 \pm 9~$MeV, while the PLSM calculations are normalized to the chiral restoration-temperature, i.e. $T_{\chi}\simeq 240\,$MeV. Also, it is obvious that the speed of sound squared ($c_s^2$) approaches the Stefan-Boltzmann limit, i.e. $1/3$, at very high temperatures. The peak, which appears near the pseudo-critical temperature, is due to the fast rate of the change of energy density with increasing temperature. 

Furthermore, the right-hand panel of Fig. \ref{fig:cs2} (b) presents the PLSM calculations of $p/\epsilon$ at  $eB=0.1$ (dotted curve), $eB=0.2$ (dotted-dash curve)  and $eB=0.3~$GeV$^2$ (double-dotted curve) and compares them to recent lattice QCD \cite{LQCD:Magnet2014} (close square), (open square) and (open triangle), respectively. A good agreement is obtained, especially at low temperature. It is obvious that the agreement is improved with increasing the magnetic field. It is noteworthy observing that the agreement looks better than that at $e B=0$.

\subsubsection{Entropy}

\begin{figure}[htb]
\centering{
\includegraphics[width=5.5cm,angle=-90]{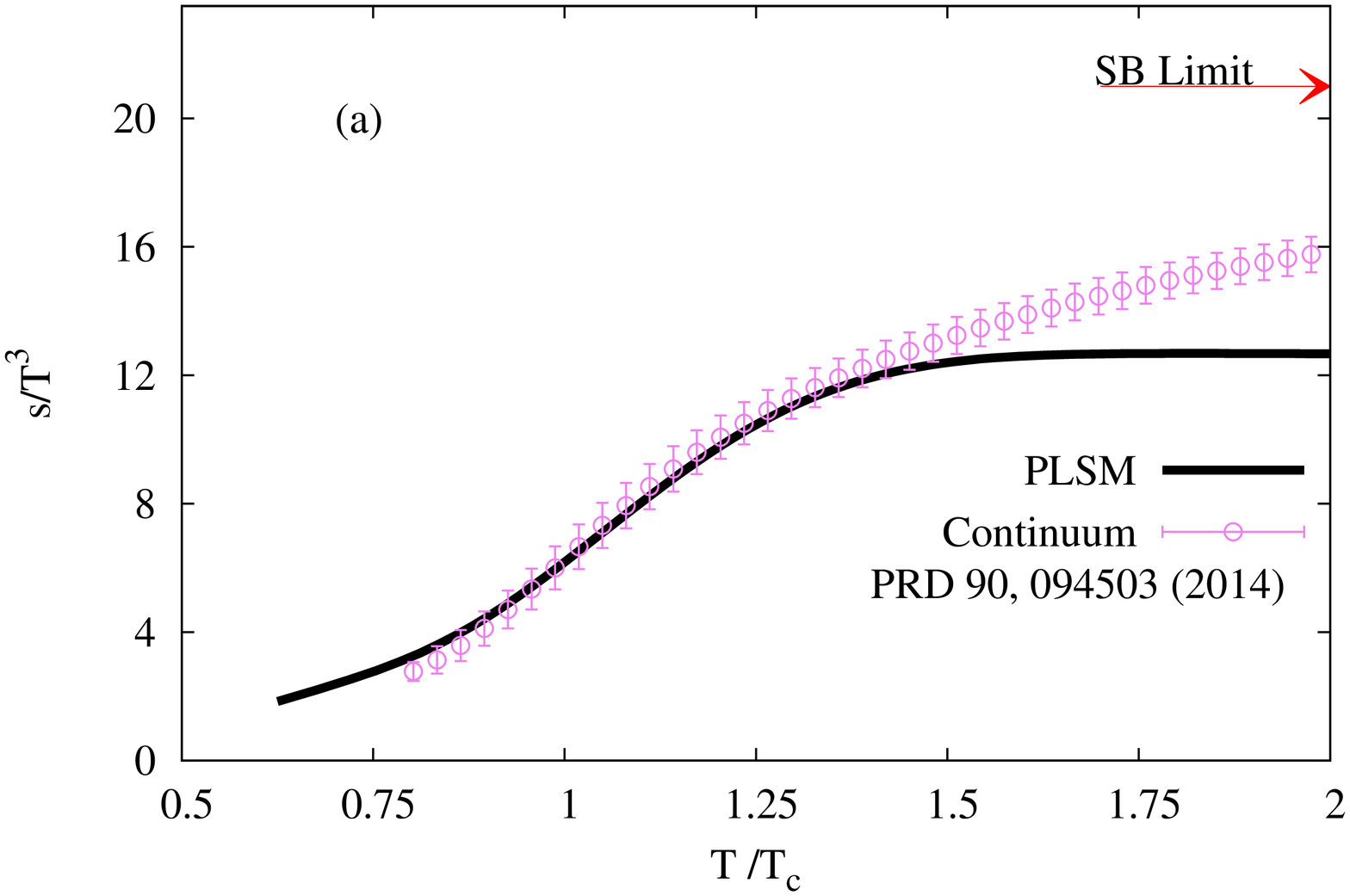}
\includegraphics[width=5.5cm,angle=-90]{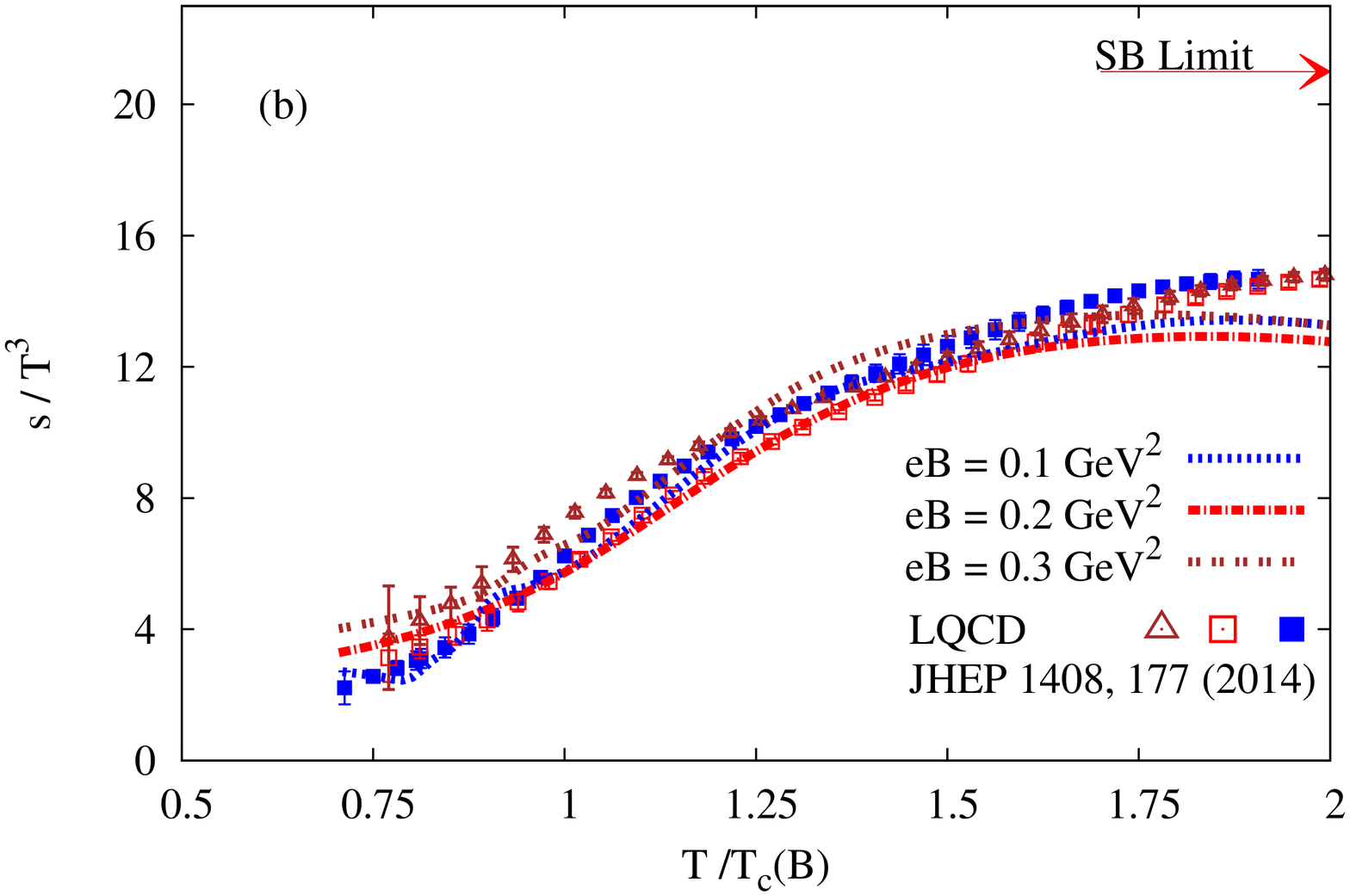}
\caption{\footnotesize Left-hand panel (a): the normalized thermal entropy density $s/T^3$ (solid curve) is depicted as a function of temperature compared with the lattice QCD calculations extrapolated to the continuum limit shown as open circles \cite{lQCD2014}. Right-hand panel (b) compares between the PLSM calculations at $eB=0.1$ (dotted curve), $eB=0.2$ (dotted-dash curve)  and $eB=0.3~$GeV$^2$ (double-dotted curve) and the recent lattice QCD (solid square), (open square) and (open triangle), respectively \cite{LQCD:Magnet2014}. 
\label{fig:scv}}}
\end{figure}  

Equation (\ref{totalfree}) expresses the explicit dependence of the free energy-density on the temperature, the chemical potential  and the magnetic field. Per definition, in B-scheme, the background magnetic field directed along $z$-axis, the thermal pressure is conjectured being isotropic. The entropy can be estimated from Eq. (\ref{eq:entropy}). Left-hand panel of Fig. \ref{fig:scv} (a) shows the normalized entropy ($s/T^3$) in a wide range of temperatures at $\mu=0$ and $e B=0$. The temperature dependence continues even above $T_c$, i.e. $s/T^3$ keeps its raise with increasing $T/T_c$. Then, it becomes slightly higher than the lattice QCD results. The dependence on the magnetic field is presented in the right-hand panel (b). The temperature dependence of the normalized entropy density ($s/T^3$) [left-hand panel (a)] is calculated from the PLSM at vanishing chemical potential and compared with recent lattice QCD simulations which are extrapolated to the continuum limit \cite{lQCD2014}, as well as with the lattice QCD with p$4$ action and $N_{\tau}=8$ (open triangles) \cite{lattice2009a} and $N_{\tau}=10$ (solid circles) \cite{lattice2009a}. A reasonable agreement with the lattice QCD simulations is obvious and the phase transitions seems to take place, smoothly. The temperature dependence continues above $T_c$, i.e. $s/T^3$ keeps increasing with increasing $T/T_c$ until it becomes slightly lower than lattice results.

\subsubsection{Specific heat}

\begin{figure}[htb]
\centering{
\includegraphics[width=5.5cm,angle=-90]{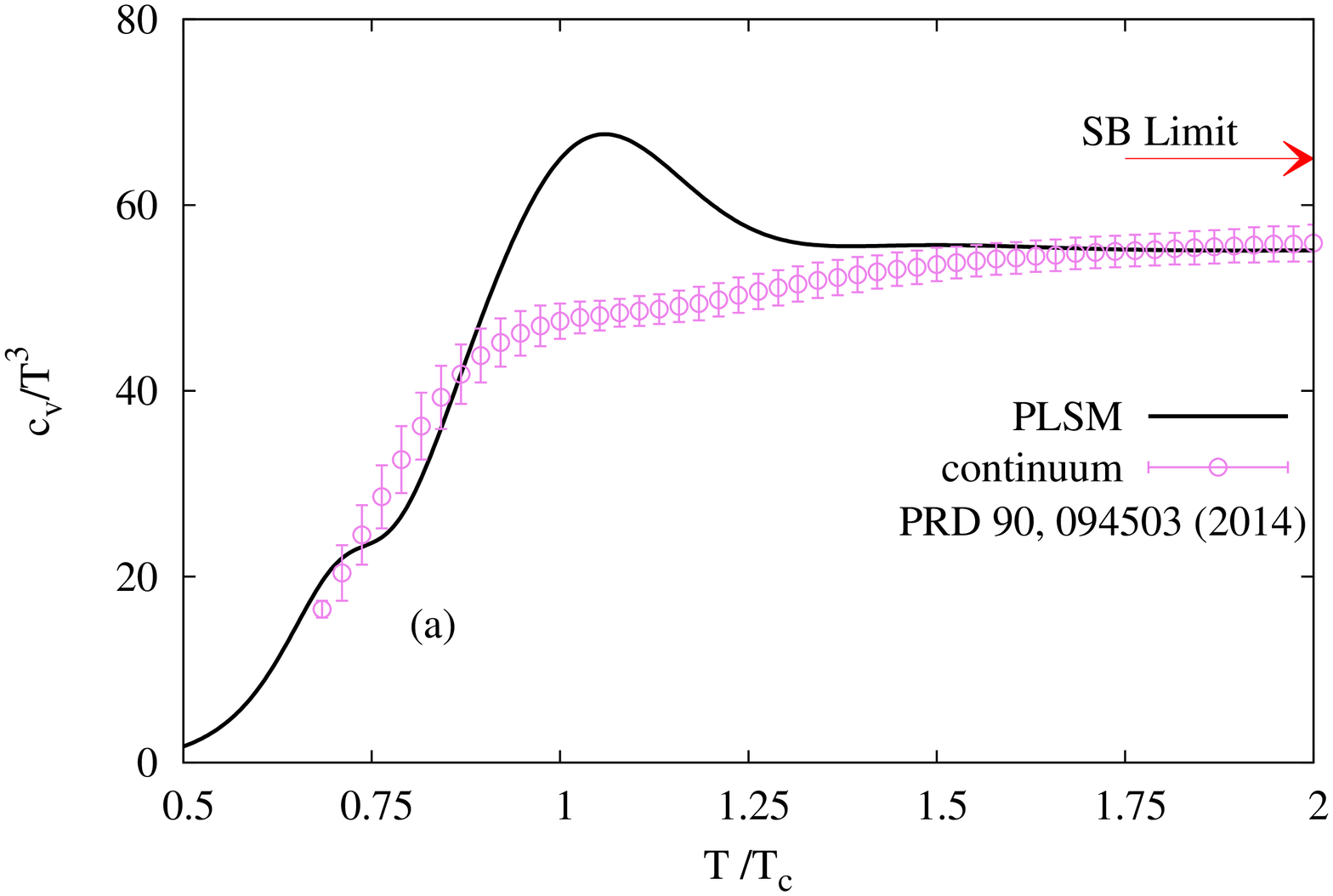}
\includegraphics[width=5.5cm,angle=-90]{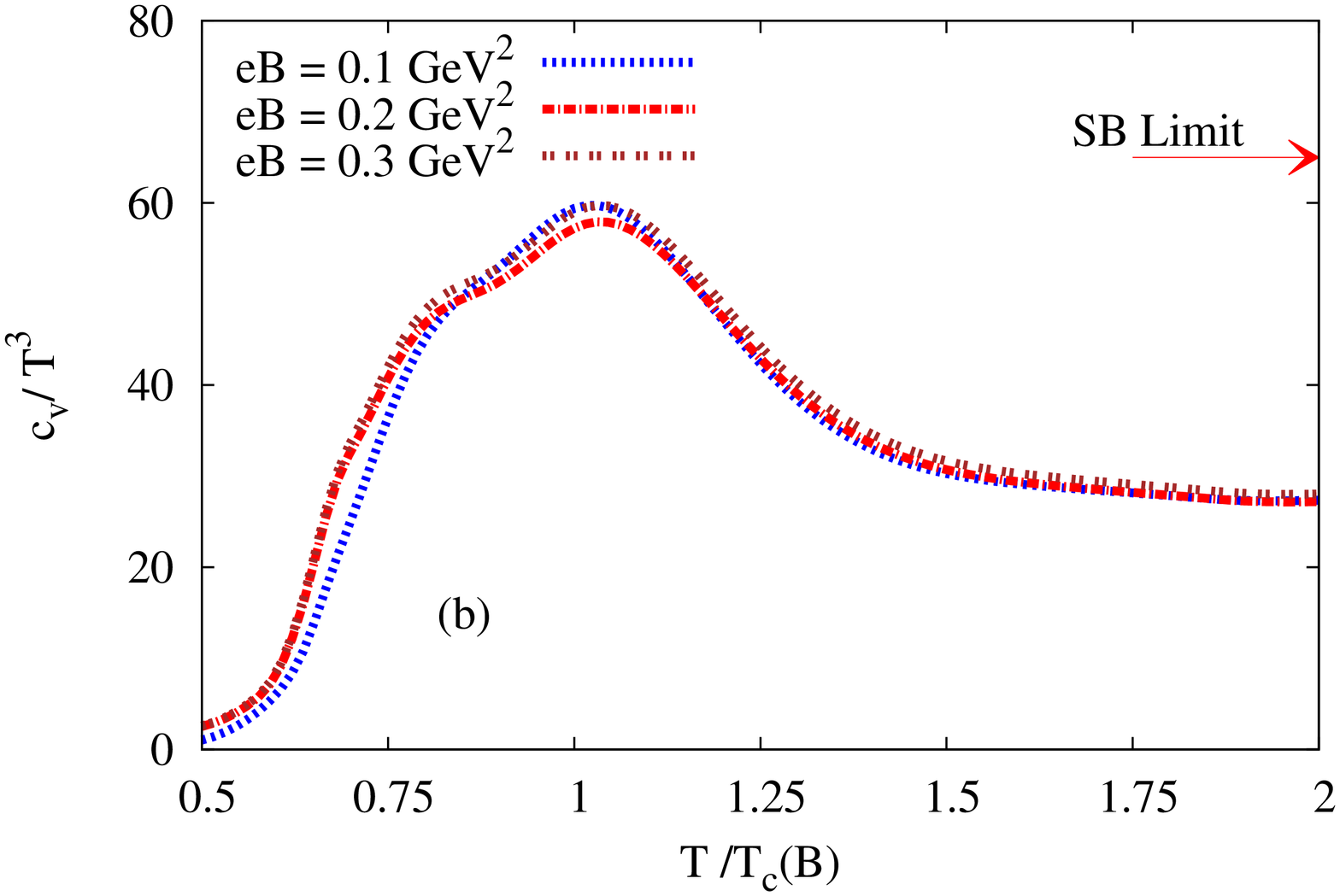}
\caption{\footnotesize Left-hand panel shows the temperature dependence of the normalized specific heat $c_v/T^3$ calculated from the PLSM at vanishing chemical potential (solid curve) and compared with the continuum limits (open circles) \cite{lQCD2014}. Right-hand panel: the influence of the magnetic field on the QCD matter is illustrated at $eB=0.0$ (solid curve),  $eB=0.1$ (dotted curve) and $eB=0.3~$GeV$^2$ (double-dotted curve). 
\label{fig:cv}}}
\end{figure}

In the previous section, we have characterized the temperature dependence of the entropy density at vanishing and finite magnetic field. Besides the entropy density, the specific heat ($c_v$) can be utilized in charactering the equation of state, i.e. speed of sound squared. 

In a wide range of temperatures and at $\mu=0$, the PLSM calculations of $c_v$ are compared with recent lattice QCD  \cite{lQCD2014} in the left-hand panel of  Fig. \ref{fig:cv} (a). The PLSM results indicate a reasonable agreement with the lattice QCD at low and high temperatures, as well. The peak positioned at $T_c$ can be interpreted from the definition of the specific heat, $c_v=\partial \epsilon/ \partial T$. Thus, the peak might be understood due to the rapid change in the energy density around $T_c$. Furthermore, the appearance of the peak would be strongly related to the chemical potential. In a previous work \cite{Tawfik:Magnetic2}, we have shown that the peak declines with increasing $\mu$. The specific heat grows with the increase in the temperature. Apart from the PLSM peak that appears at the transition temperature, the specific heat approaches the corresponding SB-limit, at high temperatures. The deviation SB limits might be due to the differences in the contributing gluon dof. The overall agreement with the lattice QCD  \cite{lQCD2014} is excellent.

The right-hand panel (b) presents normalized specific heat as derived from modified free-energy, Eq. (\ref{totalfree}), due to finite magnetization ($\mathcal{M}$) as a function of temperature at $\mu=0$ and at $eB=0.0$ (solid curve), $eB=0.1$ (dotted curve) and $eB=0.3~$GeV$^2$ (double-dotted curve). Unfortunately, there is no lattice QCD calculations exist so far to compare our PLSM results with. We notice that the deviation from SB-limit becomes relatively larger than the one observed in other thermodynamic quantities.

\subsection{Magnetic properties} 
\label{Mag_properties}

The magnetization determines the response of the system of interest to an external magnetic field, which can be produced in HIC. Because of the relativistic off-center motion of the spectators in peripheral collisions (motion of electric charge generates magnetic field perpendicular to plane of both motion direction and electric field) and because of the local imbalance in the momenta that are carried by the colliding nucleons in peripheral and even central collisions (such local imbalance leads to angular momentum and thus magnetic field) \cite{Tawfik_NICA}. Such a magnetic field can be tremendously huge, ${\cal O}(m_{\pi}^2)$, i.e. very much larger than the detector magnet field or even greater than any other magnetic field observed so far in the whole universe. 

Concretely, we shall analyse some magnetic properties of the QCD matter. This includes the magnetization and the magnetic catalysis.

\subsubsection{Magnetization}

As introduced in section \ref{thermo}, the magnetization ($\mathcal{M}$) largely affects the thermodynamic properties of strongly interacting QCD matter and can be derived from Eq. (\ref{eq:Magnet}) in GeV$^2$ in natural units. 

Sign of the magnetization manifests an important magnetic property of the system of interest. In the present case, the QCD matter is {\it para}- or {\it dia}-magnetic if $\mathcal{M}>0$ or $\mathcal{M}<0$, respectively. Furthermore, from solid state physics, one can borrow that 
\begin{itemize}
\item in {\it dia}-magnetic QCD matter, the thermal QCD medium aligns oppositely to the direction of the magnetic field and produces an induced electric current, which spreads as small loops attempting to cancel the effects of the applied magnetic field, and 
\item in {\it para}-magnetic QCD matter, the thermal QCD medium  aligns towards the direction of the magnetic field. 
\end{itemize}

\begin{figure}[htb]
\centering{
\includegraphics[width=5.5cm, angle=-90]{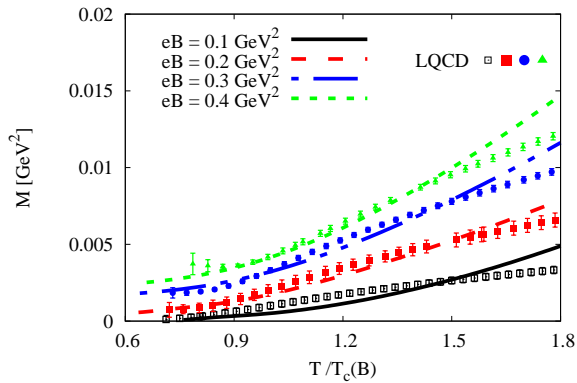}
\caption{\footnotesize The temperature dependence of the PLSM calculations on the magnetization ($\mathcal{M}$) is compared with recent lattice QCD calculations at various magnetic fields \cite{LQCD:Magnet2014}; $e B=0.1$ (solid squares and dotted curve), $e B=0.2$ (open circles and dashed curve), $e B=0.3$ (open triangles and double-dotted curve), and $e B=0.4$ (astrides and dot-dashed curve).
\label{fig:Magn_TceBA}}}
\end{figure}

Figure \ref{fig:Magn_TceBA} presents the magnetization  $\mathcal{M}$ of the QCD matter as a function of temperatures at $\mu=0$ and $eB= 0.1$ (dotted), $0.2$ (dashed), $0.3$ (double-dotted) and $0.4$ GeV$^{2}$ (dash-dotted curve). The results are compared with recent lattice QCD \cite{LQCD:Magnet2014} at $eB=0.1$ (closed square), $0.2$ (open circle), $0.3$ (open triangle) and $0.4$ GeV$^{2}$ (astrides). It is apparent that the sign of the magnetization is positive $\mathcal{M}>0$ and the resulting $\mathcal{M}$ seems to increase with increasing the magnetic field. As introduced, the positive sign of $\mathcal{M}$ indicates that the paramagnetic contribution of the QCD matter becomes dominant with increasing temperature. Within the temperature range characterizing the hadron phase (below $T_c$), the PLSM curve seems to well resemble the lattice simulations in an excellent way. At temperatures characterizing the QGP (above $T_c$), the PLSM curve becomes larger than the one representing the lattice calculations, especially at very high temperatures. In this range of temperatures, the hadrons are conjectured to deconfine into color, quark and gluon dof. It seems that these degrees of freedom are not sufficient enough to achieve a good agreement at very high temperature. 

Furthermore, the PLSM has a temperature-limited applicability depending on the temperature applicability of its order parameters, section \ref{sec:model}. Only within this limit, the capability of the PLSM to characterize the magnetic properties of  hot QCD matter is guaranteed. 

In additional to the modifications in the thermodynamic quantities due to finite magnetic field, which are mainly determined by the magnetization ($\mathcal{M}$),  it intends now to characterize other magnetic properties. As in the quantum electrodynamics (QED), the magnetic catalysis describes how the magnetic field dynamically generates masses. Furthermore, the Meissner effect describes how the magnetic field changes the order of the phase transition in type-I superconductor. These two phenomena are borrowed to study the possible influence of finite magnetic field on the QCD phase-space structure \cite{fraga2008,fraga2009} and/or the response of hadronic and partonic matter to nonzero magnetic field in thermal and dense QCD medium \cite{THD:magnetic}.

The inclusion of finite magnetic field in the QCD-effective models is achieved through modification in the phase space, energy-momentum dispersion relations and Landau levels occupations. Various works have been devoted to characterize the magnetic catalysis \cite{Ferreira2014q, Fraga:2014pte, Zhuang:2014syr, Ferreira2014s, Ferreira:2014we, Farias:2016}. Accordingly, the pseudo-critical temperatures increase as the magnetic field increases. On the other hand, other previous works have concluded a decrease in the pseudo-critical temperature as the magnetic field increases \cite{Ferreira2014s, Ferreira:2014we, Farias:2016}. In Ref. \cite{Fraga:2014pte}, it was argued that the increase in $T_c(B)$ with the magnetic field is possible unless some parameters are fine-tuned and the coupling constant is assumed to vary with the magnetic field, i.e. $g(B)$. Furthermore, it was remarked that $T_0(B)$ does not enable PLSM to reproduce $T_c(B)$ as deduced from the lattice QCD calculations, especially at magnetic fields  $eB \leq 1~$GeV$^2$. 

In the present work, it intends to assure excellent agreement between recent lattice QCD calculations and our PLSM calculations. In doing this,  at vanishing and finite magnetic field, we assume that the quark-hadron coupling constant $g=6.5$ and the Polyakov-loop deconfinement temperature $T_0 = 187~$MeV for SU($2+1$).  The main difference between this present work and other literature is the  construction of and the outcome from the PLSM free energy. In other words, the thermodynamic potential, Eq. (\ref{potential}), is the key expression distinguishing our results from all others. In this regards, there are two different mechanisms. The first one is that the magnetic field improves the phase-transition due to its contributions to the Landau quantizations and the levels. The second mechanism assumes that, the magnetic field contributes to the suppression in the chiral condensates relevant to the restoration of the chiral symmetry breaking. This suppression is known as inverse magnetic catalysis and manifests a decrease in the chiral pseudo-critical temperature with increasing magnetic field. 

When recalling the thermodynamic results, we can conclude that the temperature dependence of the thermodynamic quantities at $e B\neq0$ are smaller than those at $e B=0$. This means that the results seem to be shifted to lower,  e.g. smaller $T_c$, with increasing the magnetic field.

\subsubsection{$T$-$e B$ QCD phase diagram}

\begin{figure}[htb]
\centering{
\includegraphics[width=6.cm, angle=-90]{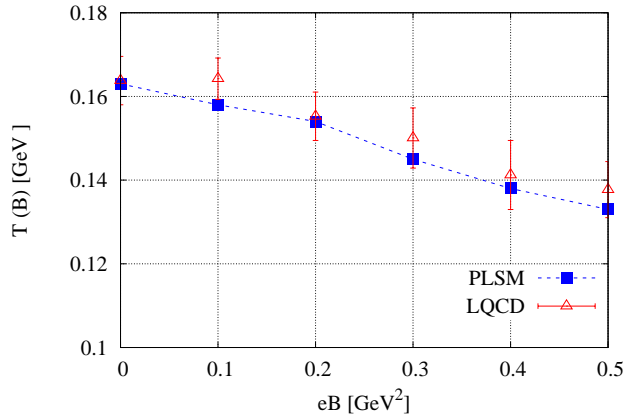}
\caption{\footnotesize The dependence of the pseudo-critical deconfinement and  chiral  temperatures on the magnetic fields as determined form PLSM (Closed symbols). The corresponding lattice QCD simulations (curved band \cite{Fodor_eB:2012} and open symbols \cite{LQCD:Magnet2014}). The vertical bands refer to averaged magnetic fields as estimated at RHIC and LHC energies. Their widths represent their relevant statistical certainties. }
\label{fig:Magn_TceBC}}
\end{figure}

Both subtracted chiral-condensates and the deconfinement order parameters at different values of magnetic field; $eB=0.0$ (solid curves), $eB=0.1$ (dotted curves), $eB=0.2$ (dashed curves), $eB=0.3$ (double-dotted curves) and $eB=0.4~$GeV$^2$ (dotted-dash curves) are used to gain a reliable estimation for their corresponding pseudo-critical temperatures. We find that the increase in the magnetic field seems to suppress the chiral condensate relevant to the restoration of the chiral symmetry breaking. Again, the earliness of the chiral-condensate suppression relative to the temperatures is due to an inverse magnetic catalysis. Thus, we find that both $T_c$ and $T_{\chi}$ decrease with increasing the magnetic field. Accordingly, we can extract the variation of the pseudo-critical temperature with increasing the magnetic field as shown in $T$-$eB$ QCD phase diagram, Fig. \ref{fig:Magn_TceBC}. 

Figure \ref{fig:Magn_TceBC} shows the variation of the pseudo-critical temperatures with the increase in the magnetic field at $\mu=0$. The PLSM results are obtained from different procedures designed for the determination of both pseudo-critical deconfinement ($T_c$) and chiral ($T_{\chi}$) temperatures. These are depicted as solid symbols. $T_c$ was estimated from the corresponding maximum point in the peak of the {\it deconfinement} susceptibility in quark number multiplicities, while, $T_\chi$ from the subtracted {\it chiral} condensate $\Delta_{l,s}$. The PLSM results are compared to recent lattice QCD simulations \cite{LQCD:Magnet2014,Fodor_eB:2012}. The curved band represents the lattice QCD simulations \cite{Fodor_eB:2012} determined from the inflection point of the strange quark-number susceptibility, while the open symbols represent the lattice QCD results \cite{LQCD:Magnet2014} as obtained from the inflection point of the normalized entropy.

It is obvious that the PLSM results are in a good agreement with both lattice QCD simulations \cite{LQCD:Magnet2014, Fodor_eB:2012}. Furthermore, we find that both PLSM and lattice QCD calculations clearly distinguish between both types of pseudo-critical temperatures, i.e. either chiral or deconfindement. Also, it is apparent that the pseudo-critical temperatures decrease with the increase in the magnetic field. This is an obvious {\it inverse} magnetic catalysis.

A few remarks on our calculations for the magnetic catalysis is now in order. The physical mechanism for the magnetic catalysis relies on a competition between the contributions of sea and valance quarks \cite{Fraga:2014pte, Endrodi:2013}. In Eq. (\ref{potential}), the first term, $U(\sigma_l,\,\sigma_s)$, refers to the contributions of valence quarks. This plays an important role in assuring spontaneous chiral symmetry breaking. The second term, $\mathbf{\mathcal{U}}(\phi, \phi^*, T)$, gives the gluonic potential contributions, while the last term, $\Omega_{\bar{q}q}(T, \mu _f, B)$ or $\Omega_{ \bar{q}q}(T, \mu _f)$, represents the contributions of sea quarks. We have noticed that a small effect from $U(\sigma_l,\,\sigma_s)$, at high temperatures. In additional to these various contributions, the influences of the Landau quantizations seem to play an essential role.

\section{Conclusions \label{Conclusion}} 

In the present work, we have combined linear-sigma model with the Polyakov-loop potentials to address chiral and deconfinement phase transitions, to characterize the quark-hadron phase structure, to determine the QCD thermodynamics at finite temperature and magnetic field, and then to estimate some of the magnetic properties. Our PLSM calculations indicate that the chiral condensates ($\sigma_l$ and $\sigma_s$) and deconfinement order parameters ($\phi$ and $\phi^*$) are sensitive to the applied magnetic field. Their agreement with recent lattice simulations is very convincing. Order parameters such as $M_b$ and $\Delta_{l,s}$ seem to have a better agreement with the lattice QCD. 

At vanishing magnetic field, we have observed a fair matching between our calculations and the lattice simulations, especially for the restoration of the broken chiral symmetry. This type of phase transition is not a rapid one but rather a {\it slow} crossover. With increasing the magnetic field, the chiral phase transition is suppressed, i.e. takes place at lower temperatures. This results in a decrease in the chiral pseudo-critical temperatures with the increase in the magnetic fields. Furthermore, we find that the chiral temperatures characterizing the light-quark condensates ($T_{\chi}^{(l)}$) decrease faster than the chiral temperatures for the strange-quark condensates ($T_{\chi}^{(s)}$). 

The PLSM at finite magnetic field assumes some restrictions to the quarks due to the existence of free charges in the plasma phase. The Yukawa coupling of quarks ($g$) seems to play an essential role in achieving an excellent agreement with the recent lattice QCD simulations. Furthermore, the QCD system undergoes modifications due to the presence of the finite magnetization. The free energy is given as an integral over the longitudinal momentum, which is directed towards the $z$-axis. Then we applied Landau theory (Landau quantization), which quantizes of the cyclotron orbits of the charged particles at finite magnetic field.  Accordingly, we find that the values of the chiral condensates are suppressed with the increase in the magnetic field. 

Furthermore, in mean field approximation, we have constructed the partition function and then estimated various thermodynamic quantities, such as the trace anomaly, the speed of sound squared, the entropy density, and the specific heat. We have studied their dependences on temperature and magnetic field in both hadron and parton phases. We have compared our calculations with recent lattice simulations at vanishing chemical potential and finite magnetic field \cite{lQCD2014}. We conclude that our calculations with a scalar coupling parameter simulate well the lattice QCD calculation at finite magnetic field  \cite{LQCD:Magnet2014}.  

At low temperatures, there is good agreement between our PLSM and lattice QCD calculations. In this temperature limit, the meson fluctuations in the mean field approximation are conjectured to be dominating. Similarly, at high temperature, the gluon dynamics seems not being fully accomplished by the Polyakov-loop potentials. This explains the discrepancy with the lattice QCD calculations. 

In point of view of Eqs. (\ref{eq100:cs2}) and (\ref{eq:cs2}), the quantities $p/\epsilon$ and $s/c_v$ should not be necessarily identical. This is only the case at vanishing chemical potential. Otherwise both quantities are different. Alternatively, it was pointed out that such a difference could be originated in the fluctuation term $\epsilon \, \partial (p/\epsilon) /\partial \epsilon $ \cite{Tawfik:2012ty}.

We conclude that increasing the magnetic field increases the thermodynamic quantities, especially in the hadron phase. At high temperatures, the PLSM thermodynamics is apparently limited below the SB limits. Furthermore, we find that the quark-hadron phase boundary is shifted to lower values of temperatures with increasing the magnetic field. 

The sign of magnetization manifests the para- or dia-magnetic properties of the QCD matter. The excellent agreement between the PLSM and recent lattice calculations means that our PLSM calculations give a well-suited description for the degrees of freedom (thermodynamics) in both hadron and parton phases. The magnetic field seems to enhance the occurrence of chiral phase-transition due to its contributions to the Landau quantizations. The magnetic field also contributes to the suppression in the chiral condensates relevant to the restoration of the chiral symmetry breaking. This suppression is known as the inverse magnetic catalysis and seems to manifest a decrease in the chiral pseudo-critical temperature with increase magnetic field.

It is noteworthy highlighting that the results estimated from PLSM, at vanishing and finite magnetic field, can be trusted within the limits of the numerical calculations of the PLSM order parameters, $\sigma_l$, $\sigma_s$, $\phi$ and $\phi^*$ in thermal and dense QCD medium. In this regard, one has to recall that the PLSM has a temperature-limited applicability depending on the temperature validity of its chiral and deconfinement order parameters. Only within this limit the conclusions drawn about the equation of state, the magnetic properties of the QCD matter and the chiral phase-structure, etc. are certain.




\begin{thebibliography}{99}


\bibitem{Kajantie:2003AS}		
K. Kajantie, M. Laine, K. Rummukainen and Y. Schr$\ddot{o}$der, Phys. Rev. D {\bf 67}, 105008 (2003). 

\bibitem{Andersen:2011QD} J. O. Andersen, L. E. Leganger, M. Strickland, and N. Su, JHEP {\bf 1108}, 053 (2011).

\bibitem{Vuorinen2016} A. Vuorinen, EPJ Web Conf. {\bf 137}, 09011 (2017). 

\bibitem{Kurkela2010GH} A. Kurkela, P. Romatschke, and A. Vuorinen, Phys. Rev. D {\bf 81}, 105021 (2010). 

\bibitem{Rummukainen10}  K. Kajantie, M. Laine, K. Rummukainen, and Y. Schr$\ddot{o}$der, Phys. Rev. D {\bf  67}, 105008 (2003).

\bibitem{Rummukainen11}  K. Kajantie, M. Laine, K. Rummukainen, and Y. Schr$\ddot{o}$der, JHEP {\bf 0304}, 036 (2003).

\bibitem{Shuryak1978}   E. V. Shuryak, Sov. Phys. JETP {\bf 47} (1978) 212 [Zh. Eksp. Teor. Fiz. {\bf 74}, 408  (1978) ]. 

\bibitem{Kapusta1979}  J. I. Kapusta, Nucl. Phys. B {\bf 148}, 461 (1979).

\bibitem{Toimela1983}  T. Toimela, Phys. Lett. B {\bf  124}, 407 (1983).

\bibitem{Arnold1995}  P. Arnold and C. X. Zhai, Phys. Rev. D {\bf 50}, 7603 (1994); Phys. Rev. D 51 (1995).

\bibitem{Freedman1978AA}  B. A. Freedman and L. D. McLerran, Phys. Rev. D {\bf 16} (1977) 1169; V. Baluni, Phys. Rev. D {\bf 17}, 2092 (1978).

\bibitem{Vuorinen2003} A. Vuorinen, Phys. Rev. D {\bf 68}, 054017 (2003).

\bibitem{Kharzeev2008} D. E. Kharzeev, L. D. McLerran, and H. J. Warringa, Nucl. Phys. A {\bf 803}, 227 (2008).

\bibitem{Skokov:2009} V. Skokov, A. Illarionov,  and  V. Toneev, 
Int. J. Mod. Phys. A {\bf 24}, 5925 (2009). 

\bibitem{Synchrotron:2010} K. Tuchin, Phys. Rev. C {\bf 82}, 034904 (2010).

\bibitem{Elec:Magnet}  A. Bzdak and V. Skokov, Phys. Lett. B {\bf 710}, 174 (2012); \\ 
W. Deng and X. Huang, Phys. Rev.  C {\bf 85}, 044907 (2012).

\bibitem{Gupta:2004} S. Gupta, Phys. Lett. B {\bf 597}, 57 (2004).

\bibitem{Bratkovskaya:1995} E. L. Bratkovskaya, O. V. teryaev, and V. D. Toneev, Phys. Lett. B {\bf 348}, 283 (1995).

\bibitem{Shovkovy:2013} I. A. Shovkovy, Lect. Notes Phys. {\bf 871}, 13 (2013).

\bibitem{Preis:2011} F. Preis, A. Rebhan, and A. Schmitt, 
JHEP {\bf 1103}, 033 (2011).

\bibitem{Sanfilippo:2010} M. D. Elia, S. Mukherjee, and F. Sanfilippo, Phys. Rev. D {\bf 82}, 051501 (2010).

\bibitem{catalysis:2014} A. Haber, F. Preis, and A. Schmitt, Phys. Rev. D {\bf 90}, 125036 (2014).

\bibitem{Catalysis:2015} M. Ferreira, P. Costa,  C. Providencia,  O. Lourenco, and T. Frederico, {\it "Inverse Magnetic Catalysis in hot quark matter within (P)NJL models0"}, 
talk at {\it ''Compact Stars in the QCD Phase Diagram IV (CSQCD IV)''}, 	26-30 Sep 2014. Prerow, Germany, 
1504.01313 [hep-ph].

\bibitem{ALICE:2012} I. Selyuzhenkov, {\it et al.}, Prog. Theor. Phys. Suppl. {\bf 193}, 153 (2012).

\bibitem{PHENIX:2014} N. N. Ajitanand, R. A. lacey, A. Taranenko, and J. M. Alexander, Phys. Rev. C {\bf 83}, 011901 (2011).

\bibitem{STAR:2009}  B. I. Abelev,  {\it et al.}, Phys. Rev. Lett. {\bf 103}, 251601 (2009).

\bibitem{STAR:2010}  B. I. Abelev, {\it et al.}, Phys. Rev. C {\bf 81}, 054908 (2010).

\bibitem{STAR:2011}  S. A. Voloshin, Indian J. Phys. {\bf 85}, 1103 (2011).

\bibitem{STAR:2014} L. Adamczyk,  {\it et al.}, Phys. Rev. Lett. {\bf 113}, 052302 (2014).

\bibitem{Fodor_eB:2012}  G. S. Bali, F. Bruckmann, G. Endrodi, Z. Fodor, S. D. Katz, S. Krieg, A. Schafer and K. K. Szabo, JHEP {\bf 1202}, 044 (2012).

\bibitem{lattice:2013b} L. Levkova and C. DeTar, Phys. Rev. Lett. {\bf 112}, 012002 (2014).

\bibitem{lattice:2013} C. Bonati, M. DElia, M. Mariti, F. Negro, and F. Sanfilippo, Phys.Rev. D {\bf 89}, 054506 (2014).

\bibitem{Endrodi:2013} F. Bruckmann, G. Endrodi, and T. G. Kovacs, JHEP {\bf 1304}, 112 (2013).

\bibitem{QCD:2013d}   G. S. Bali, F. Bruckmann, G. Endrodi, A. Schafer, PoS {\bf LATTICE2013}, 182 (2014).

\bibitem{LQCD:Magnet2014} G. S. Bali, F. Bruckmann, G. Endrodi, S.D. Katz, and A. Schafer, JHEP {\bf 1408}, 177 (2014).

\bibitem{Mizher;2010}  A. J. Mizher, M. N. Chernodub and E. S. Fraga, Phys. Rev. D {\bf 82}, 105016 (2010).

\bibitem{Skokov;2012} V. Skokov, Phys. Rev. D {\bf 85}, 034026 (2012).

\bibitem{squeezing:2013} G. Bali, F. Bruckmann, G. Endrodi, and A. Schafer, Phys. Rev. Lett. {\bf 112}, 042301 (2014).

\bibitem{HRG1} G. Endrodi, JHEP {\bf 1304}, 023 (2013).
 
\bibitem{HRG2} A. Bhattacharyya, S. K. Ghosh,  R. Ray, S. Samanta, Europhys. Lett. {\bf 115}, 62003 (2016).

\bibitem{Klevansky:1992} S. P. Klevansky, Rev. Mod. Phys.  {\bf 64}, 649 (1992).

\bibitem{NJLsu2} P. G. Allen a and N.N. Scoccola, Phys. Rev. D {\bf 88}, 094005 (2013).

\bibitem{Menezes:2009a} D.P. Menezes, M.B. Pinto, S.S. Avancini, A.P. Martinez, and C. Providencia, Phys. Rev. C  {\bf 79}, 035807 (2009).

\bibitem{Fukushima:2010l}  K. Fukushima, M. Ruggieri, and R. Gatto, Phys. Rev. D {\bf 81}, 114031 (2010).

\bibitem{Ruggieri:2013}  M. Ruggieri, M. Tachibana, and V. Greco, JHEP {\bf 1307}, 165 (2013).

\bibitem{Tawfik:LSM} A. Tawfik, N. Magdy, and A. Diab, Phys. Rev. C {\bf 89}, 055210 (2014).

\bibitem{Tawfik:Masses} A. Tawfik and  A. Diab, 
Phys. Rev. C {\bf 91}, 015204 (2015). 

\bibitem{Tawfik:Magnetic} A. Tawfik and N. Magdy, Phys. Rev. C {\bf 90}, 015204 (2014).

\bibitem{Tawfik:Magnetic2} A. Tawfik and N. Magdy, Phys. Rev. C {\bf 91}, 015206 (2015).  

\bibitem{Diab1455} Abdel Magied Abdel Aal Diab, Abdel Nasser Tawfik, M.T. Hussein, "{\it Electromagnetic Effects on Strongly Interacting QCD-Matter}", 1611.06926 [hep-lat].

\bibitem{Diab2355} Abdel Nasser Tawfik, Abdel Magied Diab and  M.T. Hussein, Int. J. Mod. Phys. A {\bf 31}, 1650175 (2016). 
 
\bibitem{Diab23} Abdel Nasser Tawfik, Abdel Magied Diab, and  M.T. Hussein,  Int. J. Adv. Res. Phys. Sci. {\bf 3}, 4 (2016).

\bibitem{review1} R. Gatto and M. Ruggieri, Lect. Notes Phys. {\bf 871}, 87 (2013).

\bibitem{review2} D. E. Kharzeev, K. Landsteiner, A. Schmitt, and H. U. Yee, Lect. Notes Phys. {\bf 871}, 1 (2013).

\bibitem{review3} Jens O. Andersen, William R. Naylor, and Anders Tranberg, 
Rev. Mod. Phys. {\bf 88}, 025001 (2016). 



\bibitem{lsm1} M. Gell-Mann and M. Levy, 
Il Nuovo Cimento {\bf 16}, 705  (1960).

\bibitem{Schwinger} J. Schwinger, Ann. Phys. (N.Y.) {\bf 2}, 407 (1957).

\bibitem{Schwinger2} J. Schwinger, Phys. Rev. {\bf 82}, 914 (1951); {\bf 91}, 713 (1953); {\bf 91}, 728 (1953); {\bf 92}, 1283 (1953); {\bf 93}, 615 (1954); {\bf 94}, 1362 (1954).

\bibitem{THD:magnetic} A. Tawfik and A. Diab, and M. T. Hussein, "{\it Chiral Magnetic Effects form Extended SU(3) Linear-Sigma Model}", in press, Annals of Physics  (2017).

\bibitem{Tawfik:quasi} A. Tawfik and N. Magdy, 
J. Phys. G {\bf 42}, 015004 (2015) . 

\bibitem{Tawfik:2013eua} A. Tawfik, 
Phys. Rev. C {\bf 88}, 035203 (2013). 

\bibitem{lQCD2014} A. Bazavov {\it et al.}, Phys. Rev. D {\bf 90}, 094503 (2014).


\bibitem{fraga2008} E. S. Fraga and A. J. Mizher,  Phys. Rev. D {\bf 78}, 025016 (2008).

\bibitem{fraga2009} E. S. Fraga and A. J. Mizher, Nucl. Phys. A {\bf 831}, 91 (2009).

\bibitem{holographicMS} A. V. Zayakin, JHEP {\bf 0807}, 116 (2008);
G. Lifschitz and M. Lippert, Phys. Rev. D {\bf 80}, 066005 (2009);\\
G. Lifschitz and M. Lippert, Phys. Rev. D {\bf 80}, 066007 (2009);\\
H. U. Yee, JHEP {\bf 0911}, 085 (2009);\\
S. Cui, Y. h. Gao, Y. Seo, S. J. Sin, and W. S. Xu, Phys. Rev. D {\bf 81}, 066001 (2010);\\
E. D’Hoker and P. Kraus, JHEP {\bf 1003}, 095 (2010).

\bibitem{Fukushima:2008}  K. Fukushima, Phys. Rev. D {\bf 77}, 114028 (2008);\\
K. Fukushima, D. E. Kharzeev, and H. J. Warringa, Phys. Rev. D {\bf 78}, 074033 (2008).

\bibitem{Klevansky:1989}  S. P. Klevansky and R. H. Lemmer, Phys. Rev. D {\bf 39}, 3478 (1989).

\bibitem{Gusynin:1995}  V. P. Gusynin, V. A. Miransky, and I. A. Shovkovy, Phys. Lett. B {\bf 349}, 477 (1995);\\
 V. P. Gusynin, V. A. Miransky, and I. A. Shovkovy, Nucl. Phys. B  {\bf 462}, 249 (1996).

\bibitem{Semenoff:1990}G. W. Semenoff, I. A. Shovkovy, and L. C. R. Wijewardhana, Phys. Rev. D  {\bf 60}, 105024 (1999).

\bibitem{Goyal:2000} A. Goyal and M. Dahiya, Phys. Rev. D {\bf 62}, 025022 (2000).

\bibitem{Hiller:2008} B. Hiller, A. A. Osipov, A. H. Blin, and J. da Providencia, SIGMA {\bf 4}, 024 (2008).

\bibitem{Rojas:2008} E. Rojas, A. Ayala, A. Bashir, and A. Raya, Phys. Rev. D  {\bf 77}, 093004 (2008).

\bibitem{Klimenko:1998} K. G. Klimenko, {\it "Magnetic catalysis and oscillating effects in Nambu-Jona-Lasinio model at nonzero chemical potential "}, 
talk at 5th Int. Workshop on {\it ''Thermal Field Theories and Their Applications''}, 10-14 Aug 1998. Regensburg, Germany, hep-ph/9809218.

\bibitem{Babansky:1998} A. Y. Babansky, E. V. Gorbar, and G. V. Shchepanyuk, Phys. Lett. B  {\bf 419}, 272 (1998).

\bibitem{Cohen:Cohen2007}  T. D. Cohen, D. A. McGady, and E. S. Werbos, Phys. Rev. C  {\bf 76}, 055201 (2007).

\bibitem{Agasian:2000}  N. O. Agasian and I. A. Shushpanov, Phys. Lett. B  {\bf 472}, 143 (2000).

\bibitem{satz2009} P. Castorina, J. Cleymans, D. E. Miller, and H. Satz, Eur. Phys. J. C {\bf 66}, 207-213 (2010).

\bibitem{Tawfik:2012ty} A. Tawfik and H. Magdy, 
Int. J. Mod. Phys. A {\bf 29}, 1450152 (2014). 

\bibitem{lqcd4} S. Borsanyi, G. Endrodi, Z. Fodor, S.D. Katz, S. Krieg, C. Ratti, and K.K. Szabo, JHEP {\bf 1208}, 053 (2012).







\bibitem{blind} O. Scavenius, A. Mocsy, I. N. Mishustin, and D. H. Rischke, 
Phys. Rev. C {\bf 64}, 045202 (2001). 


\bibitem{Weinberg1972}  S. Weinberg, {\it "Gravitation and Cosmology"}, (Wiley, New York, 1972).

\bibitem{Schaffner:2013 chiral} L. M. Haas, R. Stiele, J. Braun, J. M. Pawlowski, and J. Schaffner-Bielich, Phys. Rev. D {\bf  87}, 076004 (2013). 

\bibitem{Ratti:2005} C. Ratti, {\it et al.}, Phys. Rev. D {\bf 73}, 014019 (2005). 

\bibitem{Roessner:2007} S. Rossner, C. Ratti, and W. Weise, Phys. Rev. D {\bf 75}, 034007 (2007).

\bibitem{Schaefer:2007d}  B.-J. Schaefer, J. M. Pawlowski, and J. Wambach, Phys. Rev. D {\bf 76}, 074023 (2007).

\bibitem{Polyakov:1978vu}  A.~M.~Polyakov, 
Phys. Lett.  B {\bf 72}, 477 (1978).

\bibitem{Susskind:1979up}  L. Susskind, 
Phys. Rev.  D {\bf 20}, 2610 (1979).

\bibitem{Banks1982} T. Banks and A. Zaks, Nucl. Phys. B{\bf  196}, 189 (1982).

\bibitem{Miransky1997}  V. A. Miransky and K. Yamawaki, Phys. Rev. D{\bf 55,} 5051 (1997).

\bibitem{Appelquist1996}  T. Appelquist, J. Terning, and L. C. R. Wijewardhana, Phys. Rev. Lett. {\bf 77}, 1214 (1996).

\bibitem{Braun2006} J. Braun and H. Gies, JHEP {\bf  06}, 024 (2006).

\bibitem{Schaefer:2008hk} B.-J. Schaefer and M. ~Wagner, 
Phys. Rev. D~{\bf 79}, 014018 (2009). 

\bibitem{Schaefer:2007c} B.-J. Schaefer and J. Wambach, Phys. Rev. D {\bf 75}, 085015 (2007). 

\bibitem{Kapusta:2006pm} J. I. Kapusta and C. Gale, {\it ''Finite-temperature field theory: Principles and applications''}, 
(Cambridge University Press, Cambridge, 2006).


\bibitem{Boomsma:2010}  J. K. Boomsma and D. Boer, Phys. Rev. D {\bf 81}, 074005 (2010).

\bibitem{Fraga:2014pte} E.S. Fraga,  B.W. Mintz, and  J. Schaffner-Bielich,  Phys. Lett. B {\bf 731}, 154 (2014). 

\bibitem{Andersen2014}  Jens O. Andersen, William R. Naylor, Anders Tranberg, JHEP 1404, 187 (2014);

\bibitem{Andersen2015} Jens O. Andersen, William R. Naylor, Anders Tranberg, JHEP 1502, 042 (2015).

\bibitem{Bruckmann2013} Falk Bruckmann, Gergely Endrodi, Tamas G. Kovacs, {\it ”Inverse magnetic catalysis in QCD”}, 1311.3178 [heplat].

\bibitem{Ferreira2014q} M. Ferreira, P. Costa, D.  P. Menezes, C. Providencia, and N. Scoccola, Phys. Rev. D {\bf 89}, 016002 (2014), Addendum: Phys. Rev. D {\bf 89}, 019902 (2014).

\bibitem{Zhuang:2014syr} G. Cao, L. He, and P.  Zhuang. Phys. Rev. D {\bf 90}, 056005 (2014).

\bibitem{Ferreira2014s} M. Ferreira, P. Costa, O. Lourenço, and T. Frederico,  Phys. Rev.  D {\bf 89}, 116011 (2014). 

\bibitem{Ferreira:2014we} M. Ferreira, P. Costa, and C. Providencia,  Phys. Rev. D {\bf 90} ,016012  (2014).

\bibitem{Farias:2016} R.L.S. Farias, V.S. Timoteo, S.S. Avancini, M.B. Pinto, and G. Krein, 
Eur. Phys.J. A {\bf 53}, 101 (2017). 

\bibitem{Redlich:2010d} V. Skokov, B. Friman, E. Nakano, K. Redlich, and B.-J. Schaefer, Phys. Rev. D {\bf 82}, 034029 (2010).

\bibitem{Redlich:2010as} E. Nakano, B.-J. Schaefer, B. Stokic, B. Friman, and K. Redlich, Phys. Lett. B {\bf 682}, 401 (2010).

\bibitem{Bochkarev1986} A. Bochkarev and M. Shaposhnikov, Nucl. Phys. B{\bf 268}, 220 (1986).

\bibitem{Eletsky1993}  V. L. Eletsky, Phys. Lett. B {\bf 299}, 111 (1993).

\bibitem{Metzger:1994ss} D. Metzger, H. Meyer-Ortmanns, H.J. Pirner, Phys.Lett. B {\bf 321} 66, (1994).

\bibitem{Pirner:1994Conf} H.J. Pirner, B.J. Schaefer, {"\it QCD phase transition in hot hadronic matter"}, 
talk at Int. Symposium on Medium Energy Physics (ISMEP 94) 22-26 Aug 1994. Beijing, China,  hep-ph/9410264.

\bibitem{Narison:1989ax} S. Narison, "{\it QCD Spectral Sum Rules}", Lecture Notes on Physics, Vol. 26 (World Scientific, Singapore, 1989).


\bibitem{lattice2008a} M. Cheng {\it et al.}, Phys. Rev. D {\bf 77}, 014511 (2008).

\bibitem{lattice2009b} M. Cheng {\it et. al.}, Phys. Rev. D {\bf 81}, 054504 (2010). 

\bibitem{DGMOR1} M. Gell-Mann, R. J. Oakes, and B. Renner, Phys. Rev. {\bf 175}, 2195 (1968).

\bibitem{DGMOR2} R. F. Dashen, Phys. Rev. {\bf 183}, 1245 (1969).


\bibitem{HotQCDtree} A. Bazavov (HotQCD Collaboration) 
{\bf PoS LATTICE2011}, 182 (2011). 

\bibitem{lattice2009a} A. Bazavov {\it et. al.}, Phys. Rev. D {\bf 80}, 014504 (2009).

\bibitem{Orginos:2009} K. Orginos and D. Toussaint, Phys. Rev. D {\bf 59}, 014501 (1999);\\ 
G.P. Lepage, Phys. Rev. D {\bf 59}, 074502 (1999); \\
K. Orginos, D. Toussaint, and R. L. Sugar (MILC Collaboration), Phys. Rev. D {\bf 60}, 054503 (1999).


\bibitem{Heller:1999} U. M. Heller, F. Karsch, and B. Sturm, Phys. Rev. D {\bf 60}, 114502 (1999).

\bibitem{Peikert:1998} A. Peikert, B. Beinlich, A. Bicker, F. Karsch, and E. Laermann, Nucl. Phys. Proc. Suppl. {\bf 63}, 895 (1998). 

\bibitem{Landau:1995} L. Landau, E. Lifshitz, and L. Pitaevskii, {\it "Electrodynamics of continuous media. Course of theoretical physics"}, (Butterworth-Heinemann, 1995).

\bibitem{Piattella:2013wpa}
	Oliver F. Piattella, Julio C. Fabris, and Neven Bilic,
Class. Quant. Grav. {\bf 31}, 055006  (2014). 

\bibitem{action01}  S. Borsanyi, G. Endrodi, Z. Fodor, A. Jakovac, S. D. Katz, {\it et al.}, JHEP {\bf 1011}, 077 (2010); \\  
S. Borsanyi, Z. Fodor, C. Hoelbling, S. D. Katz, S. Krieg, {\it et al.}, Phys. Lett. B {\bf 370}, 99 (2014).

\bibitem{mgRHIC} Miklos Gyulassy and Larry McLerran, 
 Nucl. Phys. A {\bf 750}, 30 (2005). 

\bibitem{Tawfik_NICA} A. Tawfik, 
 Indian J. Phys. {\bf 91}, 93 (2017). 
 
\bibitem{cvfluct} George D. J. Phillies, {\it ''Elementary lectures in statistical mechanics''},  (Springer, New York, 2000). 

\bibitem{Tawfik:2012si} A. Tawfik, Adv. High Energy Phys. {\bf 2013},  574871 (2013). 

\end{thebibliography}
\end{document}